\documentclass[journal]{IEEEtran}

\usepackage{amsmath}
\usepackage{amssymb}
\usepackage{array}
\usepackage{booktabs}
\usepackage{cite}
\usepackage{graphicx}
\usepackage{stfloats}
\usepackage{multirow}
\usepackage{tabularx}
\usepackage{xcolor}
\usepackage{xurl}
\usepackage{tikz}
\usetikzlibrary{arrows.meta,calc,fit,positioning}
\usepackage[hidelinks]{hyperref}

\fnbelowfloat
\graphicspath{{figures/}}

\newcolumntype{Y}{>{\raggedright\arraybackslash}X}
\newcommand{\ProjectName}{\textsc{NoP Generator}}

\newcommand{\PaperFigureFontAt}[1]{\pgfmathsetmacro{\paperfiguresize}{#1*\paperfigurescale}\pgfmathsetmacro{\paperfigureleading}{1.1875*\paperfiguresize}\rmfamily\fontsize{\paperfiguresize}{\paperfigureleading}\selectfont}
\newcommand{\PaperFigureFont}{\PaperFigureFontAt{8}}
\newcommand{\ResizePaperFigure}[3]{\begingroup
  \pgfmathsetmacro{\paperfigurescale}{(#2)/(#1)}\resizebox{#1}{!}{#3}\endgroup}

\newcounter{paperalgorithm}
\newenvironment{paperalgorithm}[2]{\begin{table}\refstepcounter{paperalgorithm}\label{#1}\small\hrule\vspace{3pt}\textbf{Algorithm \thepaperalgorithm}\quad #2\par\vspace{3pt}\hrule\vspace{3pt}}{\vspace{3pt}\hrule\end{table}}

\title{\ProjectName: Package-Aware Co-Design of Chiplet Systems}

\author{Yishan~Zhang and Lei~Shen\thanks{The authors are with Fudan University, Shanghai, China
(e-mail: \mbox{\href{mailto:yishanzhang20@fudan.edu.cn}{yishanzhang20@fudan.edu.cn}};
\mbox{\href{mailto:lei_shen@fudan.edu.cn}{lei\_shen@fudan.edu.cn}}).}}

\IEEEaftertitletext{\vspace{-2\baselineskip}}

\AddToHookNext{shipout/foreground}{\put(0,-\paperheight){\raisebox{25pt}[0pt][0pt]{\makebox[\paperwidth][c]{\parbox[b]{\textwidth}{\footnotesize
          This work has been submitted to the IEEE for possible publication.\\
          Copyright may be transferred without notice, after which this version may no longer be accessible.}}}}}

\hypersetup{
  pdftitle={NoP Generator: Package-Aware Co-Design of Chiplet Systems},
  pdfauthor={Yishan Zhang, Lei Shen},
  pdfsubject={Package-aware synthesis and design-space exploration},
  pdfkeywords={chiplet, network-on-package, shared communication, cost and bandwidth, advanced packaging}
}

\begin{document}

\bstctlcite{IEEEreferenceStyle}
\maketitle

\suppressfloats[t]

\begin{abstract}
Chiplet integration enables systems beyond a single die's area limit and
creates opportunities for design reuse and cost reduction. Realizing these
benefits requires exploring chiplet boundaries, fabrication processes,
inter-chiplet networks, and packaging technologies together.
We present \ProjectName{}, a Network-on-Package (NoP) co-design framework.
Architects specify a system's functional blocks, the communication
requirements between them, and the available technologies. From these
inputs, the framework synthesizes a feasible package design by jointly
determining chiplet boundaries, process assignments, inter-chiplet topology
and routing, and the floorplan. Using the same communication requirements at
each stage, the framework explores combinations of 2.5D interconnects and 3D stacking
to minimize package cost or,
under a cost ceiling, the total bandwidth provisioned between chiplets
(installed capacity).
Our seven benchmarks model chiplet-based designs used in commercial processors
and accelerators. At a production volume of one million units, cost-oriented
search lowers geometric mean package cost by 26.9\% relative to the manual
references. With each reference's package cost as the ceiling, capacity-oriented
search meets the same communication requirements while reducing the geometric means of
installed capacity and package cost by 16.8\% and 2.1\%, respectively.
These results demonstrate the value of
automated exploration with packaging as an active design variable.
The source code and benchmarks are publicly available.
\end{abstract}

\begin{IEEEkeywords}
Advanced packaging, chiplet design, design-space exploration,
network-on-package, topology synthesis.
\end{IEEEkeywords}

\section{Introduction}
\label{sec:introduction}

\begin{figure}[t]
  \centering
  \includegraphics[width=\columnwidth]{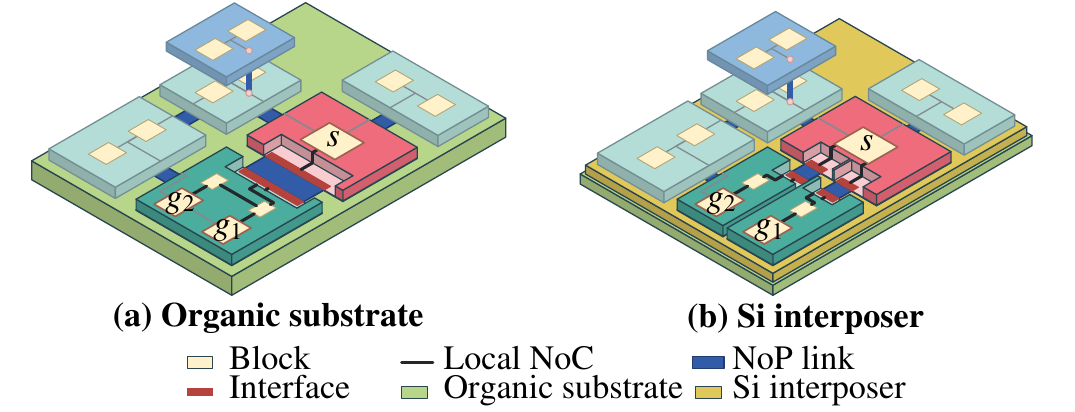}
  \caption{NoC/NoP hierarchy and packaging choices. Blocks $g_1$ and $g_2$
  share bandwidth at $s$. For the direct links shown, co-location (a) permits
  organic interconnects. With separation (b), each link must support its
  source's peak rate even if the two peaks do not coincide. The higher total
  link capacity requires denser Si interfaces to fit the available shoreline.}
  \label{fig:system-to-nop-hierarchy}
\end{figure}

\IEEEPARstart{G}{rowing} demands for computing capability and memory bandwidth
drive the integration of more compute and memory resources. The lithographic
reticle limit constrains the area of a conventional monolithic die. Chiplet
integration combines multiple dies to scale systems beyond this limit,
while providing opportunities to improve yield, reuse designs, and select
suitable processes for different functions
\cite{nvidia_blackwell_hc2024,lau2022recent,beyne20213d}.

Distributing a system across dies moves part of its communication onto
package links. Figure~\ref{fig:system-to-nop-hierarchy} shows hierarchical
packages combining organic or silicon 2.5D interconnects with 3D stacking.
Chiplet-local Networks-on-Chip (NoCs) connect functional blocks, and a
Network-on-Package (NoP) connects chiplets
\cite{yin_isca18,guirado2021wienna}.
The same blocks and communication requirements can lead to different
interface demands and packaging choices when chiplet boundaries change.
Packaging therefore affects which decompositions are attractive and which
can be physically realized.

These choices form a coupled design space. Smaller, reusable dies can
reduce manufacturing and amortized non-recurring engineering (NRE) cost,
but expose more traffic to the package. Inter-chiplet links incur area
overheads and constrain die placement and stacking.
Process assignment changes die size and cost; production
quantity changes the balance between NRE and recurring expenses.
An improvement at one design stage therefore need not translate into a
better overall design.
Manual selection covers only a small part of these alternatives. Automated
EDA support is needed to explore packaging together with partition and
inter-chiplet network design.

We present \ProjectName{} (NoPG), a co-design framework that begins with
functional blocks and communication requirements, before a chiplet
partition or inter-chiplet topology is fixed. It explores complete designs under
the available fabrication and packaging technologies. Total cost includes
chiplet manufacturing, carriers, assembly, testing, and amortized NRE.
Designers can minimize this cost or minimize installed link capacity under
a cost ceiling. The main contributions are:
\begin{itemize}
  \item \emph{Synthesis and search coordination.}
  We develop NRE-aware partitioning by batch merging, topology synthesis
  and routing under per-flow hop bounds, and cost-guided 2.5D/3D realization.
  Each stage accounts for package feasibility and bandwidth over a common
  traffic domain. A coordinator screens and schedules candidates.
  \item \emph{Design-space evaluation and insights.}
  Experiments on seven architecture-derived inputs compare synthesized
  packages with manual references and automated design baselines.
  Cost, capacity, and production-quantity studies reveal how partitioning,
  process assignment, reuse, and packaging interact, and how their
  preferred combinations change with design objectives and production volume.
\end{itemize}

The source code, benchmark inputs, and experimental scripts are
publicly available\footnote{\url{https://github.com/12zys/NoP-Generator}}
to support reproducible evaluation and further research on chiplet and NoP co-design.

\section{Background and Related Work}
\label{sec:related-work}

\subsection{Design Hierarchy and Physical Integration}

Chiplet design groups functional blocks into dies connected by a package
network~\cite{beyne20213d}. Circuit-level 3D design assigns
cells and nets to tiers for timing and physical closure, as in
LogicFolding~\cite{he2026timescaling}. A stack can itself form a chiplet
subsystem. A higher-level netlist can instantiate the entire stack as a
single chiplet and connect it through its external ports.

Chiplet networks use wired or wireless
links~\cite{yin_isca18,guirado2021wienna}.
Organic package substrates, silicon interposers, embedded bridges, and 3D stacking
differ in interconnect density, reach, yield, and assembly
requirements~\cite{lau2022recent,beyne20213d,intel_emib_brief}.
Universal Chiplet Interconnect Express (UCIe) standardizes die-to-die
interfaces~\cite{ucie30}. Planar links
need sufficient shoreline and must meet a reach limit. Vertical links
require die overlap on compatible layers.

\subsection{Chiplet Partitioning and Co-Design}

Multilevel partitioning minimizes hypergraph cut under balance
constraints~\cite{karypis1997multilevel}. Chiplet costs also depend on
each die's process~\cite{graening2026chipletpart}. Chipletizer~2.0
co-designs reusable chiplets and interfaces across products and includes
multihop latency in its objective~\cite{li2026chipletizer}.
Multi-package co-design weighs connectivity against package reliability
costs~\cite{mpcd_iccad22}. Floorplet groups modules from a synthesized
SoC hierarchy into chiplets and uses simulated traffic to guide
floorplanning~\cite{floorplet_tcad24}. CATCH models manufacturing,
test, and assembly costs~\cite{graening2026catch}.

\subsection{Network Synthesis and Communication Requirements}

Application-specific NoC synthesis couples topology, routing, and
floorplanning~\cite{murali_iccad06}. Chiplet network research
extends this work to heterogeneous interposers, modular topologies, and
resilient routing~\cite{yin_isca18,kite_dac20,feng2023scalable,red_tcad24}.
RapidChiplet estimates latency and throughput for user-specified
interconnects~\cite{iff2025rapidchiplet}.
AutoDSRI derives a shared chiplet library and reusable interposer
topology from multiple application core graphs~\cite{liu2025autodsri}.
NetSmith can bound interposer network diameter and targets uniform
random traffic in its main designs~\cite{green2024netsmith}.

NoPG enforces an inter-chiplet hop bound for each block-to-block flow
and adopts the traffic polytope model from robust
routing~\cite{benameur2005routing} to capture which flow rates can
occur together under shared bandwidth limits. As partitions
and routes change, we recompute peak aggregate loads at chiplet
boundaries and on each physical link over the same polytope. These
bandwidth requirements guide the choice of 2.5D/3D packaging
(Table~\ref{tab:related-work}).

\begin{table}[!htb]
  \caption{Inputs and decisions in chiplet co-design}
  \label{tab:related-work}
  \centering\footnotesize
  \setlength{\tabcolsep}{3pt}
  \begin{tabularx}{\columnwidth}{@{}l>{\raggedright\arraybackslash}p{0.29\columnwidth}Y@{}}
    \toprule
    Method & Key inputs & Design decisions \\
    \midrule
    ChipletPart~\cite{graening2026chipletpart}
      & Block-level netlist, process nodes
      & Chiplet partition, process assignment, 2.5D floorplan \\
    \addlinespace[2pt]
    Chipletizer 2.0~\cite{li2026chipletizer}
      & Bandwidth-weighted block graphs for a product line
      & Chiplet partition, 2.5D floorplan, die-to-die interface assignment \\
    \addlinespace[2pt]
    PlaceIT~\cite{iff2025placeit}
      & Chiplet dimensions, interface locations
      & Chiplet placement, topology \\
    \addlinespace[2pt]
    Chiplet-Gym~\cite{mishty2025chipletgym}
      & AI workload, mesh accelerator template
      & Chiplet counts, HBM placement, 2.5D/3D packaging, link settings by connection type \\
    \midrule
    NoPG
      & Functional blocks, traffic domain and hop bounds, technology libraries, production quantity
      & Chiplet partition, process assignment, routed topology, 2.5D/3D link configuration, package floorplan \\
    \bottomrule
  \end{tabularx}
\end{table}

\section{Synthesis Framework and Problem Formulation}
\label{sec:framework}
\label{sec:problem}

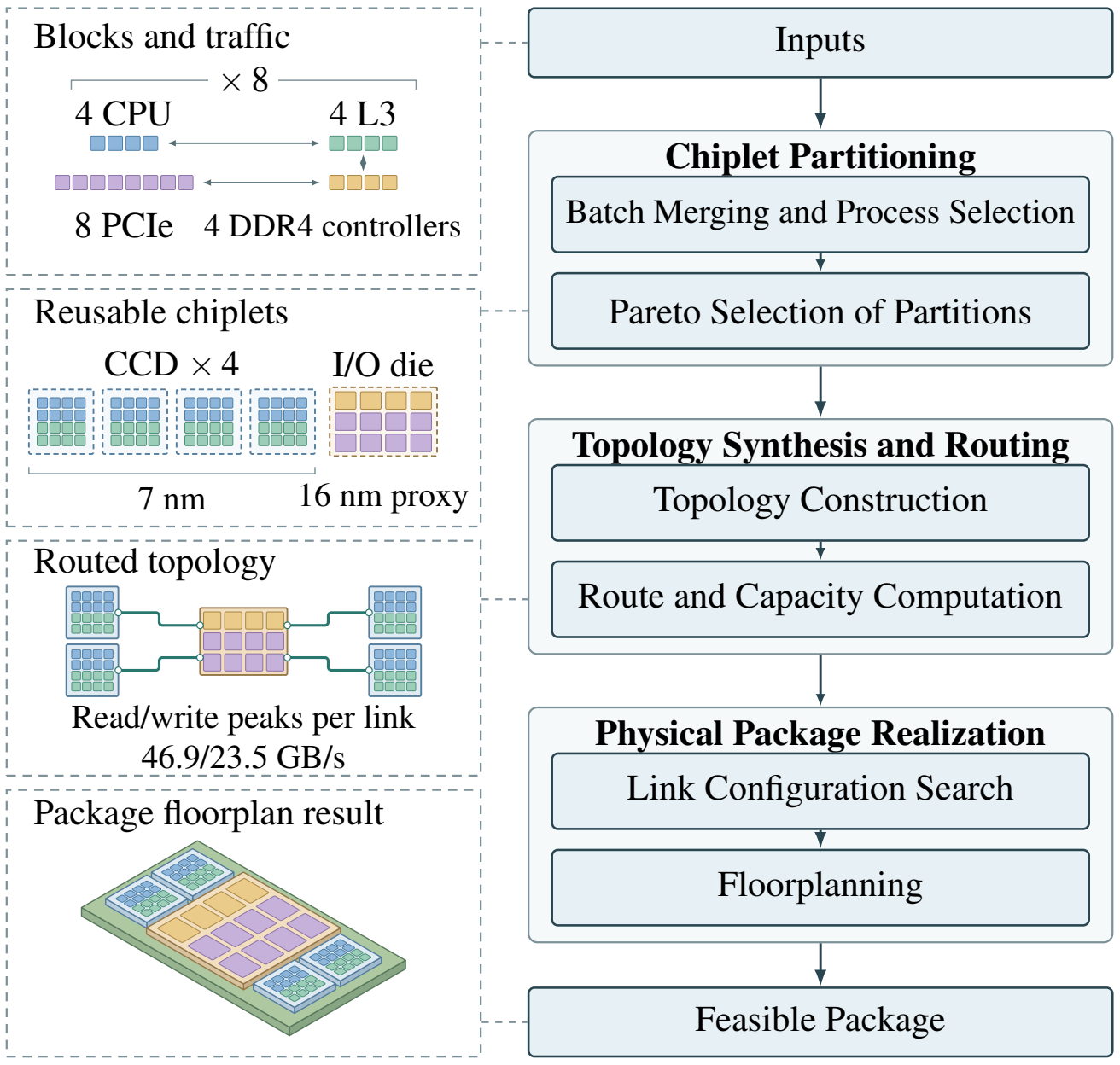
\begin{figure}
  \centering
  \ResizePaperFigure{\columnwidth}{387.61209pt}{\begin{tikzpicture}[
  x=1cm,y=1cm,font=\PaperFigureFont,
  box/.style={draw=nopInk,line width=.75pt,rounded corners=2pt,
    fill=nopFlowFill,minimum width=6.4cm,minimum height=.82cm,inner sep=0pt},
  heading/.style={font=\PaperFigureFont,text=black},
  stage/.style={draw=nopGuide,line width=.65pt,rounded corners=3pt,
    fill=nopFlowFill!42,minimum width=6.96cm,minimum height=2.80cm,inner sep=0pt},
  opbox/.style={box,minimum height=.90cm},
  stepheading/.style={font=\PaperFigureFont,text=black},
  flow/.style={-{Latex[length=2mm,width=1.45mm]},draw=nopInk,line width=.85pt},
  stepflow/.style={flow,-{Latex[length=1.5mm,width=1.2mm]}},
  back/.style={flow,draw=nopTeal,line width=.85pt},
  guide/.style={draw=nopGuide,dash pattern=on 2.4pt off 2.2pt,line width=.65pt},
  panel/.style={draw=nopGuide,line width=.65pt,dash pattern=on 3.1pt off 2.3pt,
    fill=white,minimum width=5.65cm,inner sep=0pt},
  paneltitle/.style={anchor=west,font=\PaperFigureFont,text=black},
  reftext/.style={inner sep=0pt,text=black,font=\PaperFigureFont},
  dieheading/.style={reftext,font=\PaperFigureFont},
  traffic/.style={{Latex[length=1.05mm,width=.8mm]}-{Latex[length=1.05mm,width=.8mm]},
    draw=nopInk!78,line width=.52pt},
  bracket/.style={draw=nopGuide,line width=.43pt},
  diecase/.style={inner sep=0pt,outer sep=0pt,rounded corners=.6pt,line width=.55pt},
  copycase/.style={diecase,dash pattern=on 1.8pt off 1.2pt},
  wire/.style={draw=nopTeal,line width=.9pt,line cap=round,rounded corners=1.1pt},
  pin/.style={circle,fill=white,draw=nopTeal,line width=.35pt,inner sep=0pt,minimum size=2.2pt},
  tile/.style={line width=.28pt,rounded corners=.42pt,line join=round},
  core tile/.style={tile,fill=coreBlue,draw=coreEdge},
  cache tile/.style={tile,fill=cacheGreen,draw=cacheEdge},
  ddr tile/.style={tile,fill=ddrGold,draw=ddrEdge},
  pcie tile/.style={tile,fill=pciePurple,draw=pcieEdge}
]
  \definecolor{nopInk}{HTML}{294650}
  \definecolor{nopTeal}{HTML}{28776F}
  \definecolor{nopGuide}{HTML}{7D9199}
  \definecolor{nopFlowFill}{HTML}{E7F0F4}
  \definecolor{coreBlue}{HTML}{8EB6DA}
  \definecolor{coreEdge}{HTML}{497DA6}
  \definecolor{cacheGreen}{HTML}{9BCDB9}
  \definecolor{cacheEdge}{HTML}{4C9278}
  \definecolor{ddrGold}{HTML}{EACB89}
  \definecolor{ddrEdge}{HTML}{AA8642}
  \definecolor{pciePurple}{HTML}{C5B1D9}
  \definecolor{pcieEdge}{HTML}{8D70A8}
  \definecolor{ccdFace}{HTML}{DFEAF3}
  \definecolor{ccdEdge}{HTML}{547D9E}
  \definecolor{ioFace}{HTML}{F1DFC0}
  \definecolor{ioEdge}{HTML}{957748}
  \definecolor{carrierFace}{HTML}{AEC9A2}
  \definecolor{carrierEdge}{HTML}{59795D}
  \path[use as bounding box] (0,0) rectangle (13.5,-12.80);

  \def\epycTiles#1#2#3{\foreach \col in {1,...,#1}
      \foreach \row in {1,...,#2}
        \path[#3] ({\col-(#1+1)/2-.41},{\row-(#2+1)/2-.41})
          -- ++(.82,0) -- ++(0,.82) -- ++(-.82,0) -- cycle;
  }
  \def\epycCCD{\begin{scope}[shift={(0,1)}]\epycTiles{4}{2}{core tile}\end{scope}
    \begin{scope}[shift={(0,-1)}]\epycTiles{4}{2}{cache tile}\end{scope}
  }
  \def\epycIO#1#2{\begin{scope}[x={({#1/4},0)},y={(0,{#2/3})}]
      \begin{scope}[shift={(0,1)}]\epycTiles{4}{1}{ddr tile}\end{scope}
      \begin{scope}[shift={(0,-.5)}]\epycTiles{4}{2}{pcie tile}\end{scope}
    \end{scope}
  }

  \node[panel,minimum height=3.18cm] (inputpanel) at (2.875,-1.74) {};
  \node[paneltitle] at (.25,-.48) {Blocks and traffic};
  \draw[bracket] (.82,-1.12)--(.82,-1.01)--(4.93,-1.01)--(4.93,-1.12);
  \node[reftext,fill=white,inner xsep=3pt] at (2.875,-1.01) {$\times$\,8};
  \node[reftext] at (1.45,-1.40) {4 CPU};
  \node[reftext] at (4.30,-1.40) {4 L3};
  \node[reftext] at (1.45,-2.74) {8 PCIe};
  \node[reftext,anchor=east,font=\PaperFigureFontAt{7.2}] at (5.48,-2.74) {4 DDR4 controllers};
  \foreach \x/\y/\cols/\kind in
    {1.45/-1.76/4/core tile,4.30/-1.76/4/cache tile,
     1.45/-2.23/8/pcie tile,4.30/-2.23/4/ddr tile} {
    \begin{scope}[shift={(\x,\y)},scale=.21]\epycTiles{\cols}{1}{\kind}\end{scope}
  }
  \draw[traffic] (1.96,-1.76)--(3.79,-1.76);
  \draw[traffic] (4.30,-1.89)--(4.30,-2.10);
  \draw[traffic] (2.38,-2.23)--(3.79,-2.23);

  \node[panel,minimum height=2.83cm] (partitionpanel) at (2.875,-4.915) {};
  \node[paneltitle] at (.25,-3.81) {Reusable chiplets};
  \node[dieheading] at (2.02,-4.40) {CCD $\times$\,4};
  \node[dieheading] at (4.54,-4.40) {I/O die};
  \foreach \x in {.70,1.58,2.46,3.34} {
    \node[copycase,draw=ccdEdge,fill=ccdFace!28,minimum size=.77cm] at (\x,-5.08) {};
    \begin{scope}[shift={(\x,-5.08)},scale=.15]\epycCCD\end{scope}
  }
  \node[copycase,draw=ioEdge,fill=ioFace!28,minimum width=1.28cm,minimum height=.82cm]
    at (4.54,-5.08) {};
  \begin{scope}[shift={(4.54,-5.08)}]\epycIO{1.20}{.76}\end{scope}
  \draw[bracket] (.315,-5.62)--(.315,-5.70)--(3.725,-5.70)--(3.725,-5.62);
  \node[reftext,font=\PaperFigureFontAt{7.4}] at (2.02,-5.99) {7 nm};
  \node[reftext,font=\PaperFigureFontAt{7.4}] at (4.54,-5.99) {16 nm proxy};

  \node[panel,minimum height=2.80cm] (networkpanel) at (2.875,-7.90) {};
  \node[paneltitle] at (.25,-6.78) {Routed topology};
  \begin{scope}[shift={(2.875,-7.70)},scale=.80]
    \node[diecase,draw=ioEdge,fill=ioFace,minimum width=1.04cm,minimum height=.80cm]
      (iod) at (0,0) {};
    \epycIO{1.25}{.9375}
    \foreach \dieX/\dieSide/\ioSide in {-2.255/east/west,2.255/west/east} {
      \foreach \dieY/\portY in {.43/.24,-.43/-.24} {
        \node[diecase,draw=ccdEdge,fill=ccdFace,minimum size=.62cm] (ccd) at (\dieX,\dieY) {};
        \begin{scope}[shift={(ccd.center)},scale=.161]\epycCCD\end{scope}
        \coordinate (ioport) at (iod.\ioSide |- 0,\portY);
        \coordinate (bend) at ($(ccd.\dieSide)!.5!(ioport)$);
        \draw[wire] (ccd.\dieSide)--(bend |- ccd.\dieSide) |- (ioport);
        \node[pin] at (ccd.\dieSide) {};
        \node[pin] at (ioport) {};
      }
    }
  \end{scope}
  \node[reftext,align=center,font=\PaperFigureFontAt{7.4}] at (2.875,-8.84)
    {Read/write peaks per link\\46.9/23.5 GB/s};

  \node[panel,minimum height=3.18cm] (packagepanel) at (2.875,-11.06) {};
  \node[paneltitle] at (.25,-9.78) {Package floorplan result};
  \begin{scope}[shift={(3.42,-12.40)},x={(.073cm,.042cm)},y={(-.073cm,.042cm)},
    z={(0cm,.13cm)},scale=.78,line join=round,line width=.42pt]
    \def\epycSlab#1#2#3#4#5#6#7{\path[draw=#7,fill=#6!55!#7] (#1,#2,0)--(#3,#2,0)--(#3,#2,#5)--(#1,#2,#5)--cycle;
      \path[draw=#7,fill=#6!73!#7] (#1,#2,0)--(#1,#4,0)--(#1,#4,#5)--(#1,#2,#5)--cycle;
      \path[draw=#7,fill=#6] (#1,#2,#5)--(#3,#2,#5)--(#3,#4,#5)--(#1,#4,#5)--cycle;
    }
    \epycSlab{0}{0}{24.310}{43.514}{1}{carrierFace}{carrierEdge}
    \begin{scope}[shift={(0,0,1)}]
      \foreach \x/\y/\w/\face/\edge/\mosaic in
        {12.655/32.912/8.602/ccdFace/ccdEdge/\epycCCD,3.052/32.912/8.602/ccdFace/ccdEdge/\epycCCD,
         2.000/11.602/20.310/ioFace/ioEdge/{\epycIO{4.7}{4.7}},
         12.655/2.000/8.602/ccdFace/ccdEdge/\epycCCD,3.052/2.000/8.602/ccdFace/ccdEdge/\epycCCD} {
        \epycSlab{\x}{\y}{\x+\w}{\y+\w}{.7}{\face}{\edge}
        \begin{scope}[shift={({\x+\w/2},{\y+\w/2},.7)},scale={.20*\w}]\mosaic\end{scope}
      }
    \end{scope}
  \end{scope}

  \coordinate (rightaxis) at (9.75,0);
  \node[box,heading,minimum width=6.96cm,anchor=north]
    (input) at (inputpanel.north -| rightaxis) {Inputs};
  \node[box,heading,minimum width=6.96cm,anchor=south]
    (result) at (packagepanel.south -| rightaxis) {Feasible Package};

  \node[stage,anchor=north] (partition) at ([yshift=-.615cm]input.south) {};
  \node[heading] at ([yshift=-.35cm]partition.north) {\textbf{Chiplet Partitioning}};
  \node[opbox,stepheading,font=\PaperFigureFontAt{7.5}] (merge) at ([yshift=-1.01cm]partition.north)
    {Batch Merging and Process Selection};
  \node[opbox,stepheading,anchor=south] (select) at ([yshift=.20cm]partition.south)
    {Pareto Selection of Partitions};

  \node[stage,anchor=north] (network) at ([yshift=-.615cm]partition.south) {};
  \node[heading] at ([yshift=-.35cm]network.north) {\textbf{Topology Synthesis and Routing}};
  \node[opbox,stepheading] (construct) at ([yshift=-1.01cm]network.north)
    {Topology Construction};
  \node[opbox,stepheading,anchor=south]
    (route) at ([yshift=.20cm]network.south)
    {Route and Capacity Computation};

  \node[stage,anchor=north]
    (package) at ([yshift=-.615cm]network.south) {};
  \node[heading] at ([yshift=-.35cm]package.north) {\textbf{Physical Package Realization}};
  \node[opbox,stepheading] (options) at ([yshift=-1.01cm]package.north)
    {Link Configuration Search};
  \node[opbox,stepheading,anchor=south] (floorplan) at ([yshift=.20cm]package.south)
    {Floorplanning};

  \draw[flow] (input.south)--(partition.north);
  \draw[flow] (partition.south)--(network.north);
  \draw[stepflow] (network.south)--(package.north);
  \draw[stepflow] (merge.south)--(select.north);
  \draw[stepflow] (construct.south)--(route.north);
  \draw[flow] (options.south)--(floorplan.north);
  \draw[flow] (package.south)--(result.north);

  \draw[guide] (inputpanel.east |- input.west)--(input.west);
  \draw[guide] (partitionpanel.east |- select.west)--(partition.west |- select.west);
  \draw[guide] (networkpanel.east |- route.west)--(network.west |- route.west);
  \draw[guide] (packagepanel.east |- result.west)--(result.west);
\end{tikzpicture}}
  \caption{Synthesis flow (right). The left column illustrates the inputs
  and stage outputs using the AMD EPYC~7452~\cite{amd2020epyc,lenovo2023rome}
  as the reference design.}
  \label{fig:intent-to-package}
\end{figure}

\begin{figure*}
  \centering
  \ResizePaperFigure{\textwidth}{512.14938pt}{\input{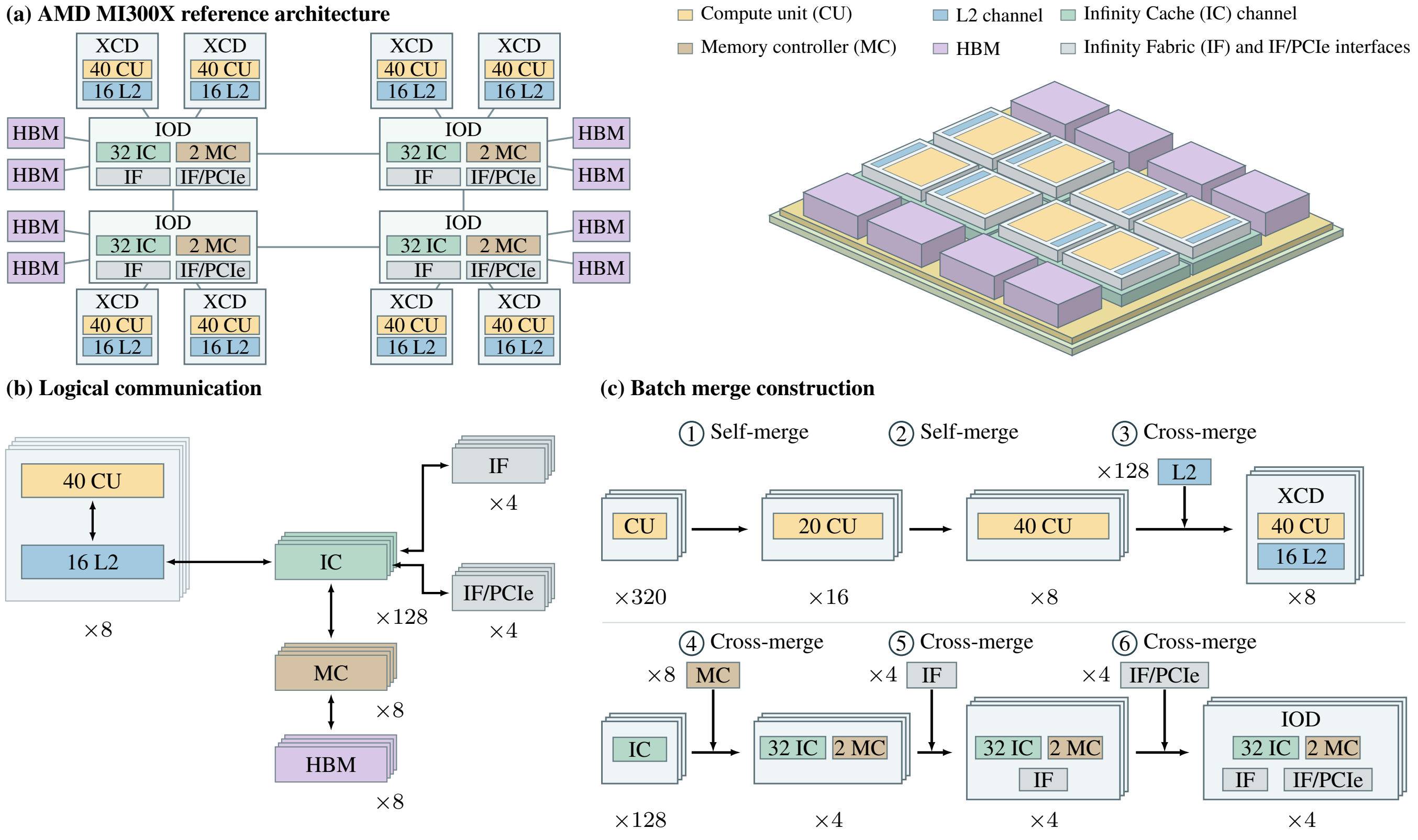}}
  \caption{Model of the major functional blocks publicly documented for the
  AMD MI300X~\cite{smith2024mi300x,amd2025cdna3}.
  (a) Reference architecture, with accelerator complex dies (XCDs) stacked
  on I/O dies (IODs) above a Si interposer. (b) Logical flows subject
  to shared bandwidth limits. (c) Illustrative batch merges
  forming eight XCD and four IOD groups. Each numbered step submits a
  complete batch. High-bandwidth memory (HBM) stacks remain fixed.}
  \label{fig:mi300x-search-example}
\end{figure*}

\subsection{Overall Synthesis Framework}

NoPG takes typed functional blocks, communication requirements,
process and interface libraries, packaging technologies, and production
quantity as inputs (Fig.~\ref{fig:intent-to-package}). Partitioning
produces a pool of candidate partitions, each with a process assignment
for its reusable die designs. For each candidate explored further, we
synthesize an inter-chiplet topology, route the logical flows, and compute
the required capacity of each link. Package realization selects a link
configuration (interfaces, packaging technologies, and die starting layers),
determines interface counts, and checks floorplan feasibility. Retaining
multiple partitions lets the search base its final choice on the feasible
packages it finds (Section~\ref{sec:coordination}).

Each chiplet is modeled as a NoP node whose local NoC is assumed to support
both intra-chiplet communication and transit forwarding.
Table~\ref{tab:notation} summarizes the principal notation.

\begin{table}
  \caption{Principal Notation}
  \label{tab:notation}
  \centering
  \footnotesize
  \setlength{\tabcolsep}{3pt}
  \begin{tabularx}{\columnwidth}{@{}lY@{}}
    \toprule
    Symbol & Meaning \\
    \midrule
    $\mathcal B,\mathcal F$ & Blocks and directed logical flows \\
    $s_f,t_f$ & Source/destination blocks of flow $f$ \\
    $x_f$ & Traffic rate of flow $f$ \\
    $\Omega$ & Admissible traffic domain \\
    $\sigma_\Omega(\boldsymbol a)$ & Worst-case weighted aggregate bandwidth over $\Omega$ \\
    $H_f$ & NoP hop bound for flow $f$ \\
    $\pi$ & Block-to-chiplet map \\
    $G=(V_\pi,E)$ & Undirected chiplet topology \\
    $\mathcal G_\pi,n_g$ & Reuse classes; instance count of class $g$ \\
    $\boldsymbol\tau,\tau_g$ & Processes of all reuse classes / class $g$ \\
    $B_{\mathrm{cut}}(\pi)$ & Total boundary-bandwidth score of partition $\pi$ \\
    $P(\pi)$ & Count of undirected communicating chiplet pairs \\
    $c_e$ & Required capacity of edge $e$ \\
    $B_{\mathrm{req}}(G),B_{\mathrm{pkg}}(d)$ & Total required / installed link capacity, each edge once \\
    $d,Q$ & Design; production quantity \\
    $D$ & Total die cost, including amortized NRE \\
    $C$ & Package cost on the final floorplan \\
    \bottomrule
  \end{tabularx}
\end{table}

\subsection{Communication Model}

Functional blocks \(\mathcal B\) exchange directed logical flows
\(f\in\mathcal F\). Each block has a reference area and allowed process
domain. Flow \(f\) has rate \(x_f\geq0\), source block \(s_f\in\mathcal B\),
and destination block \(t_f\in\mathcal B\). Flows with the same source and
destination remain distinct; reverse traffic is represented explicitly.

Each path carrying a nonzero fraction of \(f\) may traverse at most
\(H_f\) inter-chiplet links. When \(H_f=0\), the endpoints must share
a chiplet.
The traffic domain \(\Omega\) specifies which flow rates can occur
together. With \(\boldsymbol x=(x_f)_{f\in\mathcal F}\), we define
\begin{equation}
 \begin{aligned}
 \Omega&=\{\boldsymbol x=\boldsymbol T\boldsymbol z:
       \boldsymbol z\geq0,\ \boldsymbol A\boldsymbol z\leq\boldsymbol b\},\\
 &\boldsymbol T\geq0,\qquad \boldsymbol A\geq0,\qquad \boldsymbol b\geq0 .
 \end{aligned}
 \label{eq:traffic-domain}
\end{equation}
The nonnegative rate variables \(\boldsymbol z\) describe admissible
activity. The matrix \(\boldsymbol T\) maps these rates to logical flows:
one rate may contribute to several flows, and one flow may contain several
rates. Each row of \(\boldsymbol A\) describes resource use, and the
corresponding entry of \(\boldsymbol b\) gives its budget. These constraints
may overlap and span multiple blocks and flows. All coefficients
and bounds are finite, and each rate variable occurs with a positive
coefficient in at least one resource row, making the domain bounded.
This representation supports nonnegative linear traffic mappings and
upper-budget constraints, with fixed proportions encoded in
\(\boldsymbol T\).

For one 40-CU/16-L2 group in our MI300X model
(Fig.~\ref{fig:mi300x-search-example}(b)), let \(z_i^r,z_i^w\) be CU
\(i\)'s total read/write rates in GB/s.
Our MI300X input distributes each rate equally among the 16 L2 channels
through \(\boldsymbol T\). With L2 capacities from~\cite{amd2025cdna3},
we impose
\begin{equation}
 \begin{gathered}
 z_i^r+z_i^w\leq268.8\quad\forall i,\\
 \frac{1}{16}\sum_i z_i^r\leq268.8,\qquad
 \frac{1}{16}\sum_i z_i^w\leq134.4.
 \end{gathered}
 \label{eq:mi300x-cu-l2-domain}
\end{equation}
For one L2 channel, the 40 individual read peaks sum to
\(40\times16.8=672\)~GB/s. Their joint peak under \(\Omega\) is
268.8~GB/s. Sixteen reading and eight other writing
CUs at full rate yield 6.45~TB/s per XCD (51.6~TB/s across eight XCDs).
The complete input also includes IC access and a shared IC-client budget.

To evaluate outgoing bandwidth at a chiplet boundary, set \(a_f=1\) for
flows leaving the chiplet and \(a_f=0\) otherwise. For one link direction,
\(a_f\) is the fraction of flow \(f\) routed through it. With
\(\boldsymbol a=(a_f)_{f\in\mathcal F}\), the worst-case aggregate bandwidth is
\begin{equation}
 \sigma_\Omega(\boldsymbol a)=
 \max_{\boldsymbol x\in\Omega}\sum_{f\in\mathcal F}a_f x_f .
 \label{eq:traffic-support}
\end{equation}
We evaluate every partition over the same traffic domain \(\Omega\).
A flow between blocks on the same chiplet stays on die and counts once
toward that chiplet's local NoC load.
We use~\eqref{eq:traffic-support} to compute the peak aggregate load
and~\eqref{eq:local-noc-area} to estimate the corresponding local-NoC area.

\subsection{Design Variables, Constraints, and Objectives}

Let \(d\) denote a complete design with partition \(\pi\), process
assignment \(\boldsymbol\tau\), and topology \(G=(V_\pi,E)\). The vertices
are chiplets, and the edges are their direct connections. Routing fractions
\(\boldsymbol\alpha\) determine the required edge capacities \(c_e\) over
\(\Omega\). Package realization selects a link configuration,
an assembly order, and a package floorplan, and determines interface counts.
A feasible design
must serve every traffic scenario in \(\Omega\) while satisfying the
declared hop bounds, adjacency requirements, interface constraints, and
configured connection caps. Process assignments must be
admissible; floorplans must satisfy non-overlap, shoreline, reach,
vertical-overlap, reticle, and carrier limits.

At production quantity \(Q\), \(C(d,Q)\) is the cost per package on
the final floorplan, including manufacturing, testing, assembly, yield
losses, and amortized NRE.
Let \(B_{\mathrm{pkg}}(d)\) be the total installed link capacity, counting
each link once. Over the set \(\mathcal D\) of feasible designs, the
cost and capacity objectives are, respectively,
\begin{align}
 &\min_{d\in\mathcal D} C(d,Q), \label{eq:cost-objective}\\
 &\min_{d\in\mathcal D} B_{\mathrm{pkg}}(d)
 \quad\text{s.t.}\quad C(d,Q)\leq C_{\max}. \label{eq:budget-bandwidth}
\end{align}
A reference design evaluated at the same \(Q\) can supply \(C_{\max}\).
Section~\ref{sec:package-cost} details the cost model.

\subsection{Candidate Evaluation and Search Coordination}
\label{sec:coordination}

\begin{figure}
  \centering
  \ResizePaperFigure{\columnwidth}{255.30699pt}{\begin{tikzpicture}[
  x=1cm,y=1cm,
  heading/.style={text=black,font=\PaperFigureFont\bfseries,
    inner sep=0pt,align=center},
  note/.style={text=black,font=\PaperFigureFont,
    inner sep=0pt,align=center},
  queuerank/.style={circle,draw=sfCandidate,line width=.65pt,fill=white,
    text=black,font=\PaperFigureFont\bfseries,
    inner sep=0pt,minimum size=8.4pt},
  exitnote/.style={note,font=\PaperFigureFontAt{7.2}},
  flow/.style={draw=sfInk,line width=.65pt,line cap=round,line join=round,
    -{Latex[length=1.45mm,width=1.05mm]}},
  feedback/.style={draw=sfFeedback,line width=.65pt,line cap=round,line join=round,
    dash pattern=on 2.2pt off 1.8pt,rounded corners=3pt},
  feedarrow/.style={feedback,-{Latex[length=1.4mm,width=1.0mm]}},
  unit/.pic={
    \path[draw=sfGuide,line width=.45pt,fill=sfSide]
      (.49,-.43)--(.54,-.39)--(.54,-.30)--(.49,-.34)--cycle;
    \path[draw=sfGuide,line width=.45pt,fill=sfLight]
      (.22,-.34)--(.27,-.30)--(.54,-.30)--(.49,-.34)--cycle;
    \path[draw=sfGuide,line width=.5pt,fill=sfMachine]
      (.22,-.43) rectangle (.49,-.34);
    \foreach \x in {.275,.435}{
      \filldraw[draw=sfInk!80,line width=.3pt,fill=white] (\x,-.385) circle (.021);
    }
    \path[draw=sfGuide,line width=.45pt,fill=sfSide]
      (.40,-.34)--(.47,-.29)--(.47,.80)--(.40,.75)--cycle;
    \path[draw=sfInk!80,line width=.5pt,fill=sfLight]
      (.30,-.34) rectangle (.40,.75);
    \draw[draw=sfGuide!55,line width=.35pt] (.332,-.29)--(.332,.71);
    \path[draw=sfGuide,line width=.45pt,fill=sfMachine]
      (.14,.64)--(.30,.48)--(.30,.64)--cycle;
    \path[draw=sfInk!80,line width=.5pt,fill=sfSide]
      (.40,.64)--(.47,.695)--(.47,.855)--(.40,.80)--cycle;
    \path[draw=sfInk!80,line width=.5pt,fill=sfLight]
      (-.29,.80)--(-.22,.855)--(.47,.855)--(.40,.80)--cycle;
    \path[draw=sfInk!80,line width=.55pt,fill=sfMachine]
      (-.29,.64) rectangle (.40,.80);
    \foreach \x in {.19,.25,.31}{
      \draw[draw=sfGuide,line width=.45pt] (\x,.69)--(\x,.75);
    }
    \path[draw=sfGuide,line width=.45pt,fill=sfSide]
      (.08,.855)--(.12,.89)--(.12,.965)--(.08,.93)--cycle;
    \path[draw=sfInk!80,line width=.45pt,fill=sfMachine]
      (-.15,.855) rectangle (.08,.93);
    \path[draw=sfGuide,line width=.4pt,fill=sfLight]
      (-.15,.93)--(-.11,.965)--(.12,.965)--(.08,.93)--cycle;
    \foreach \x in {-.17,.17}{
      \draw[draw=sfGuide,line width=.95pt] (\x,.36)--(\x,.64);
      \draw[draw=white,line width=.3pt] ({\x-.007},.37)--({\x-.007},.63);
    }
    \path[draw=sfInk!80,line width=.5pt,fill=sfMachine,rounded corners=.6pt]
      (-.105,.47) rectangle (.105,.64);
    \draw[draw=sfGuide,line width=.4pt] (-.085,.60)--(.085,.60);
    \path[draw=sfGuide,line width=.4pt,fill=sfLight]
      (-.033,.36) rectangle (.033,.47);
    \path[draw=sfGuide,line width=.4pt,fill=sfSide]
      (.21,.27)--(.255,.305)--(.255,.395)--(.21,.36)--cycle;
    \path[draw=sfGuide,line width=.4pt,fill=sfLight]
      (-.21,.36)--(-.165,.395)--(.255,.395)--(.21,.36)--cycle;
    \path[draw=sfInk!80,line width=.5pt,fill=sfMachine]
      (-.21,.27) rectangle (.21,.36);
    \path[draw=sfFeedback!80!sfInk,line width=.45pt,fill=sfTool]
      (-.145,.205) rectangle (.145,.27);
    \path[draw=sfGuide,line width=.45pt,fill=sfLight]
      (.31,.37) rectangle (.48,.54);
    \draw[draw=sfInk!75,line width=.65pt] (.345,.49)--(.445,.49);
    \fill[sfInk!85] (.36,.415) circle (.019);
    \filldraw[draw=sfGuide,line width=.3pt,fill=white] (.435,.415) circle (.019);
  },
  screen/.pic={
    \path[draw=sfScreen!75,line width=.4pt,fill=sfScreenSide]
      (.155,-.19)--(.198,-.166)--(.198,.334)--(.155,.31)--cycle;
    \path[draw=sfScreen,line width=.7pt,fill=sfScreenFill]
      (-.035,-.30)--(.155,-.19)--(.155,.31)--(-.035,.20)--cycle;
    \foreach \v in {-.20,-.10,0,.10}{
      \draw[draw=sfScreen!75,line width=.4pt] (-.035,\v)--(.155,{\v+.11});
    }
    \foreach \u in {.028,.091}{
      \draw[draw=sfScreen!75,line width=.4pt]
        (\u,{-.30+(\u+.035)*.11/.19})--(\u,{.20+(\u+.035)*.11/.19});
    }
  }
]
  \definecolor{sfInk}{HTML}{242424}
  \definecolor{sfGuide}{HTML}{8093A2}
  \definecolor{sfGray}{HTML}{AAAAAA}
  \definecolor{sfCandidate}{HTML}{316B96}
  \definecolor{sfFeedback}{HTML}{242424}
  \definecolor{sfMachine}{HTML}{D5E2EC}
  \definecolor{sfLight}{HTML}{F1F6FA}
  \definecolor{sfSide}{HTML}{98B0C2}
  \definecolor{sfBelt}{HTML}{F2F5F7}
  \definecolor{sfTool}{HTML}{333333}
  \definecolor{sfScreen}{HTML}{777777}
  \definecolor{sfScreenSide}{HTML}{C5C5C5}
  \definecolor{sfScreenFill}{HTML}{F1F1F1}
  \path[use as bounding box] (-.08,1.67) rectangle (8.77,-3.91);

  \draw[feedback] (.78,-2.50)--(-.025,-2.50)--(-.025,1.48)--(8.04,1.48);
  \draw[feedarrow] (3.28,1.48)--(3.28,.33);
  \draw[feedarrow] (8.04,1.48)--(8.04,.33);
  \draw[feedarrow] (-.025,-1.05)--(5.35,-1.05)--(5.35,-2.17);
  \node[note,fill=white,inner xsep=4pt] at (5.42,1.48) {Pruning feedback};

  \pic[scale=.68] at (.92,0) {unit};
  \foreach \x in {4.94,5.68,6.42}{\pic[scale=.68] at (\x,0) {unit};}
  \foreach \x in {7.16,6.42}{\pic[xscale=-.68,yscale=.68] at (\x,-2.50) {unit};}
  \foreach \x in {4.09,3.52,2.95,2.38,1.81}{
    \pic[xscale=-.68,yscale=.68] at (\x,-2.50) {unit};
  }
  \node[heading] at (1.64,1.00) {Partition exploration};
  \node[note,font=\PaperFigureFont,text=black]
    at (2.16,.39) {Candidate pool};
  \node[heading] at (5.65,1.00) {Topology synthesis and routing};
  \node[heading] at (6.90,-1.50) {Link configuration};
  \node[heading] at (2.82,-1.50) {Floorplanning};

  \def\sfConveyor{
    (.92,0)--(8.03,0) .. controls (8.75,0) and (8.75,-2.50)
    .. (8.03,-2.50)--(.78,-2.50)
  }
  \draw[draw=sfGuide!60,line width=10pt,line cap=round,line join=round] \sfConveyor;
  \draw[draw=sfBelt,line width=8.7pt,line cap=round,line join=round] \sfConveyor;
  \foreach \y/\lasttick in {0/19,-2.50/20}{
    \foreach \i in {0,...,\lasttick}{
      \pgfmathsetmacro{\x}{.97+.35*\i}
      \draw[draw=sfGuide!50,line width=.35pt] (\x,{\y-.13})--({\x+.055},{\y-.165});
    }
  }

  \foreach \x/\y/\direction in {
    1.19/0/1,3.20/0/1,6.98/0/-1,7.96/0/-1,
    5.92/-2.50/1,5.27/-2.50/-1,1.27/-2.50/1}{
    \begin{scope}[shift={(\x,\y)},xscale=\direction]
      \draw[draw=sfGray!55,line width=4.3pt,line cap=round]
        (0,-.035) .. controls (.08,-.25) and (.17,-.45) .. (.40,-.45)--(.66,-.45);
      \draw[draw=sfGray!10,line width=3.3pt,line cap=round]
        (0,-.035) .. controls (.08,-.25) and (.17,-.45) .. (.40,-.45)--(.66,-.45);
      \foreach \px/\py in {.24/-.39,.57/-.45}{
        \filldraw[draw=white,line width=.4pt,fill=sfGray!85] (\px,\py) circle (.060);
      }
    \end{scope}
  }
  \node[exitnote] at (1.68,-.69) {Invalid};
  \node[exitnote] at (3.92,-.69) {Screened out};
  \node[exitnote] at (6.28,-.69) {No result};
  \node[exitnote] at (7.61,-.69) {Screened out};
  \node[exitnote] at (1.80,-3.19) {No result};
  \node[exitnote] at (4.71,-3.19) {Screened out};
  \node[exitnote] at (6.45,-3.19) {No result};

  \foreach \x/\y/\direction in {2.73/0/1,7.52/0/1,
    7.75/-2.50/-1,5.70/-2.50/-1,1.09/-2.50/-1}{
    \draw[flow] ({\x-\direction*.12},\y)--({\x+\direction*.14},\y);
  }
  \foreach \x/\y in {1.92/0,2.16/0,2.40/0,
    4.94/0,5.68/0,6.42/0,6.70/0,
    7.16/-2.50,6.42/-2.50,6.10/-2.50,3.52/-2.50,
    2.95/-2.50,2.38/-2.50,1.81/-2.50}{
    \filldraw[draw=white,line width=.55pt,fill=sfCandidate] (\x,\y) circle (.073);
  }

  \foreach \x/\y in {3.20/0,7.96/0,5.27/-2.50}{\pic at (\x,\y) {screen};}
  \foreach \t/\rank/\topx/\bottomx in {.25/3/3.65/4.97,.50/2/4.00/4.63,.75/1/4.35/4.29}{
    \pgfmathsetmacro{\bendx}{8.03+3*.72*\t*(1-\t)}
    \pgfmathsetmacro{\bendy}{-2.50*(3*\t*\t-2*\t*\t*\t)}
    \ifnum\rank=1
      \tikzset{thisrank/.style={queuerank,fill=sfCandidate,text=white}}
    \else
      \tikzset{thisrank/.style={queuerank}}
    \fi
    \node[thisrank] at (\topx,0) {\rank};
    \node[thisrank] at (\bendx,\bendy) {\rank};
    \node[thisrank] at (\bottomx,-2.50) {\rank};
  }

  \begin{scope}[shift={(.78,-2.50)}]
    \filldraw[draw=white,line width=.7pt,fill=sfCandidate] (0,0) circle (.16);
    \draw[white,line width=.85pt,line cap=round,line join=round]
      (-.075,0)--(-.02,-.055)--(.088,.06);
  \end{scope}
  \node[heading,font=\PaperFigureFont] at (.75,-1.91) {Feasible\\packages};
  \fill[sfCandidate] (.32,-3.70) circle (.073);
  \node[note,anchor=west] at (.49,-3.70) {Candidate};
  \fill[sfGray!85] (2.43,-3.70) circle (.073);
  \node[note,anchor=west] at (2.60,-3.70) {Discarded};
  \pic[scale=.63] at (4.49,-3.70) {screen};
  \node[note,anchor=west] at (4.74,-3.70) {Screen \& rank};
  \node[queuerank,fill=sfCandidate,text=white] at (7.11,-3.70) {1};
  \node[note,anchor=west] at (7.34,-3.70) {Next};
\end{tikzpicture}}
  \caption{Candidate screening, ranking, and pruning feedback from feasible
  packages. Machine counts indicate relative evaluation time per candidate.}
  \label{fig:candidate-search-funnel}
\end{figure}
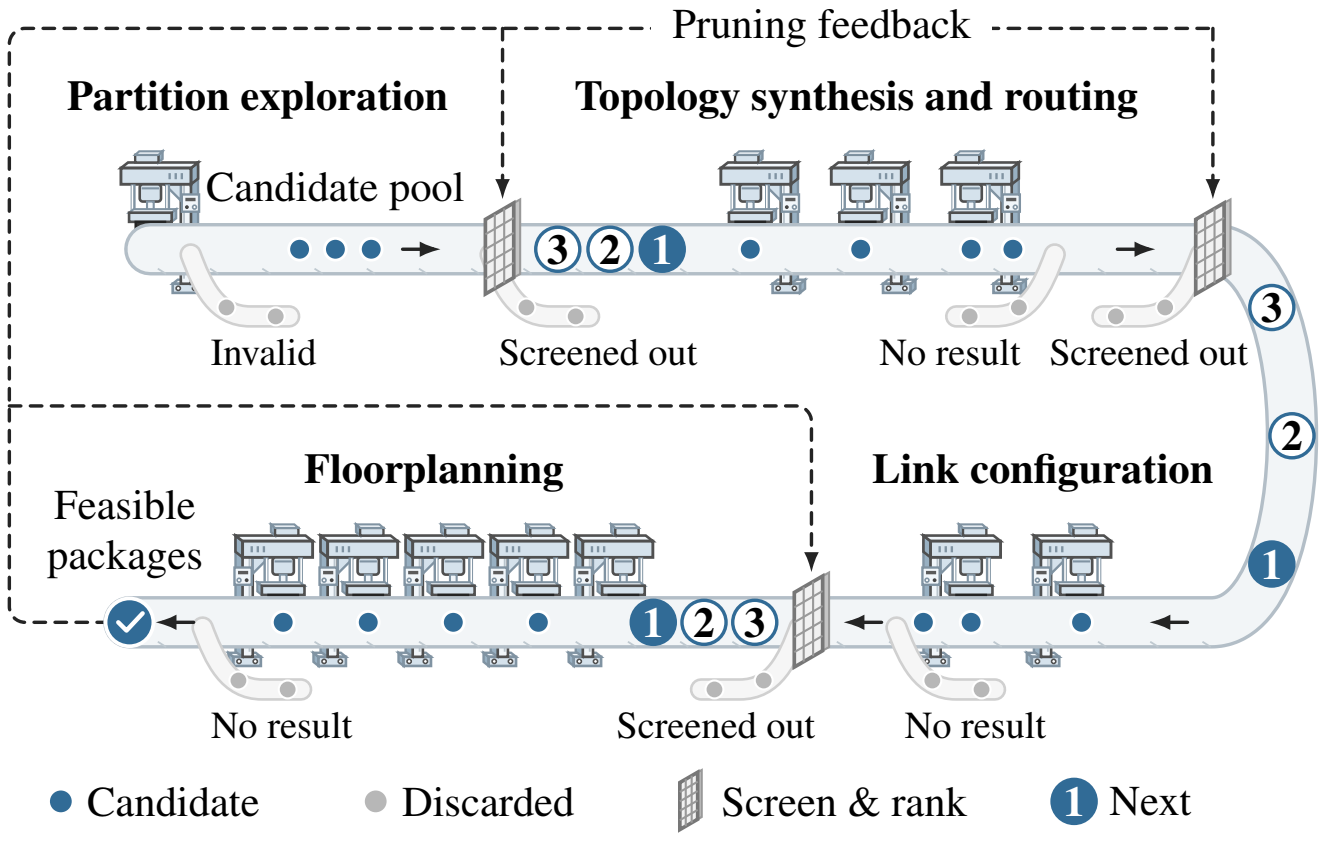

After partition exploration produces a candidate pool, candidates enter a
pipeline of topology synthesis and routing, link configuration, and floorplanning
(Fig.~\ref{fig:candidate-search-funnel}). Each candidate follows these
stages in order. The coordinator assigns ready tasks to a shared worker
pool, so several floorplan searches can run in parallel while other
candidates are still being routed or configured.

For each exported partition, the coordinator uses die costs and unavoidable
interface requirements under \(\Omega\) to screen unfinished searches.
Within the search budget, each admitted partition yields at most one routed
topology and one complete link configuration. The coordinator screens this
configuration by cost and installed capacity before floorplanning.

A design can update the incumbent once its floorplan and shared
interface template are feasible and it meets any prescribed cost limit.
The incumbent sets the cost threshold in cost mode. In capacity mode,
installed capacity sets the threshold and cost breaks ties.

\section{Chiplet Partitioning}
\label{sec:stage1}

Partitioning determines die composition, design reuse, process choices,
and the communication exposed to the package. Algorithm~\ref{alg:partition}
explores these coupled choices through batch merges of existing dies.
It evaluates the resulting partitions for die cost and communication,
retaining tradeoffs for the downstream stages.

\subsection{Search Overview and Partition Evaluation}
\label{sec:partition-objectives}

A search state \(s=(\pi,\boldsymbol\tau)\) specifies a chiplet partition
and its process assignment. The initial partition \(\pi_0\) applies
necessary co-location and leaves other blocks separate. Fixed input dies
remain exclusive.
For each state selected for expansion, Algorithm~\ref{alg:partition}
evaluates the partitions produced by batch merges and adds a representative
subset of the admitted candidates to the search pool
(Section~\ref{sec:partition-search}).
Fig.~\ref{fig:mi300x-search-example}(c) illustrates this process through
batch merges forming the reference XCD and IOD groups. After each numbered
step~\mbox{(1--6)}, the resulting partition is evaluated and selected for
further exploration.
We evaluate each state by
\(\boldsymbol q(s)=(D(s),B_{\mathrm{cut}}(\pi),P(\pi))\),
capturing die cost, boundary bandwidth, and connectivity scale,
respectively. Pareto selection retains tradeoffs among these scores for
downstream synthesis.

Chiplets with the same block-type composition form a reuse class \(g\in\mathcal G_\pi\)
whose \(n_g=|g|\) instances share one die design and process \(\tau_g\).
As in Chipletizer~2.0~\cite{li2026chipletizer}, we amortize the design's
NRE over its \(n_g Q\) copies across \(Q\) packages. The die cost per
package is
\begin{equation}
 D(s)=\frac{N_{\mathrm{IP}}}{Q}+\sum_{g\in\mathcal G_\pi}
 \left[\frac{N_g(\tau_g)}{Q}+n_g C_g^{\mathrm{rec}}(\tau_g)\right],
 \label{eq:stage1-die-ledger}
\end{equation}
where \(N_{\mathrm{IP}}\) counts one development fee per distinct input
block type. Section~\ref{sec:package-cost} details the NRE per die
design \(N_g\) and the recurring cost per copy \(C_g^{\mathrm{rec}}\).

For scalable logic processes, a linear bandwidth--area model~\cite{wang2024tmac}
gives the local-NoC area per die in class~\(g\) as
\begin{equation}
 A_{\mathrm{NoC},g}=\kappa_{\mathrm{NoC}}(\tau_g)
 \max_{u\in g}B_{\mathrm{NoC},u},
 \label{eq:local-noc-area}
\end{equation}
where \(B_{\mathrm{NoC},u}\) is the worst-case aggregate bandwidth over
\(\Omega\) of all flows with at least one endpoint in die \(u\); flows
entirely within \(u\) are counted once.
The area-per-bandwidth coefficient \(\kappa_{\mathrm{NoC}}(\tau_g)\)
is estimated from router-area data~\cite{zhu2025theseus} with
logic-process scaling. This common area enters each instance's
die cost, so equal block compositions can have different costs under
different mappings.

Under partition map \(\pi:\mathcal B\rightarrow V_\pi\), flow \(f\)
connects dies \(\pi(s_f)\) and \(\pi(t_f)\), crossing a die boundary when
they differ. All resource constraints in \(\Omega\) remain in force during
aggregation. For boundary scoring, we group crossing flows by whether
the two directions use independent (\(\mathrm I\)) or shared
(\(\mathrm S\)) transmission resources. General links use a common mode
specified before synthesis. Only standalone fixed blocks, such as HBM
stacks, may have prescribed interfaces. Links with such requirements use
the specified mode and carry only the corresponding block's direct traffic.
For a cut \(A\subseteq V_\pi\) and mode
\(\mu\in\{\mathrm I,\mathrm S\}\), let \(F_\pi^\mu(A)\) and
\(R_\pi^\mu(A)\) be the supports of the outgoing and incoming crossing
traffic, computed by~\eqref{eq:traffic-support}. The bandwidth score
\(K_\pi(A)\) of the cut and its sum over single-die cuts are
\begin{equation}
 \begin{aligned}
 K_\pi(A)={}&2\max\{F_\pi^{\mathrm I}(A),R_\pi^{\mathrm I}(A)\}\\
            &+\max\{F_\pi^{\mathrm S}(A),R_\pi^{\mathrm S}(A)\},\\
 B_{\mathrm{cut}}(\pi)={}&\sum_{u\in V_\pi}K_\pi(\{u\}).
 \end{aligned}
 \label{eq:strict-cut}
\end{equation}
Independent transmit/receive resources provision both directions at the
larger directional peak. For shared resources, the larger directional
peak gives a capacity lower bound. Each link provides capacity at both
endpoints, so these per-die bounds yield \(B_{\mathrm{cut}}(\pi)/2\leq
B_{\mathrm{pkg}}(d)\) for any feasible realization~\(d\)~of~\(\pi\).

Let \(\mathcal E_\pi^{\mathrm{comm}}\) contain the unordered pairs of
distinct parts with declared communication. The third score is
\begin{equation}
 P(\pi)=|\mathcal E_\pi^{\mathrm{comm}}|.
 \label{eq:partition-pair-count}
\end{equation}
Multiple logical flows and opposite directions count as one pair. These
pairs need not become direct physical edges: routing may connect them
through other chiplets, subject to hop and connection limits.
\(P\) complements \(B_{\mathrm{cut}}\): partitions with similar boundary
bandwidth can involve different numbers of communicating chiplet pairs.

\subsection{Batch Merging and Process Selection}
\label{sec:partition-merges}

Each action merges groups of existing dies without splitting them.
A batch action evaluates design replacement before intermediate NRE
overheads affect search selection. For example,
\[
 4\mathrm A+4\mathrm B
 \longrightarrow3\mathrm A+3\mathrm B+\mathrm{AB}
 \longrightarrow4\mathrm{AB}.
\]
At fixed processes, the die-specific NRE term changes from
\((N_{\mathrm A}+N_{\mathrm B})/Q\) to
\((N_{\mathrm A}+N_{\mathrm B}+N_{\mathrm{AB}})/Q\), then to
\(N_{\mathrm{AB}}/Q\). Old designs stop incurring NRE only when their
last instances disappear. The search therefore combines fine binary
steps with batch actions at these inventory boundaries.

For current reuse classes \(\mathrm A,\mathrm B\), a recipe makes
\(m\) copies of \(r_A\mathrm A+r_B\mathrm B\). A self-merge uses
\(r_A\geq2,r_B=0\); a cross-merge uses \(r_A,r_B\geq1\).
With inventories \(n_A,n_B\), the maximum count is
\begin{equation}
 m_{\max}=
 \begin{cases}
 \lfloor n_A/r_A\rfloor, & r_B=0,\\
 \min\{\lfloor n_A/r_A\rfloor,\lfloor n_B/r_B\rfloor\}, & r_B>0 .
 \end{cases}
 \label{eq:partition-merge-recipe}
\end{equation}
For binary recipes (\(r_A+r_B=2\)), all counts \(1,\ldots,m_{\max}\) are
candidates. For higher arities, only \(m=m_{\max}\) is submitted, and
at least one participating old class must disappear:
\(mr_A=n_A\), or \(mr_B=n_B\) for a cross-merge. All self-retirement and
joint-retirement cross recipes are retained. To reduce the remaining
cross-recipe count, single-sided retirement requires at least one of
\(r_A,r_B\) to be 1 or prime.

Communication structure guides the choice of instances within each recipe.
Let \(\pi'\) be the partition after merging \(u,v\) into \(w\).
The fixed traffic domain and removal of internal flows give
\begin{equation}
 K_{\pi'}(\{w\})=K_\pi(\{u,v\})
 \leq K_\pi(\{u\})+K_\pi(\{v\}).
 \label{eq:merge-cut-subadditivity}
\end{equation}
The merged die's boundary score cannot exceed the sum of the original
scores; other dies' scores remain unchanged. Hence
\(\Delta B_{\mathrm{cut}}=B_{\mathrm{cut}}(\pi')-B_{\mathrm{cut}}(\pi)\leq0\).
Writing \(\mathcal N_\pi(u)\) for the communication neighbors of \(u\),
\begin{equation}
 \Delta P=-\mathbf1[\{u,v\}\in\mathcal E_\pi^{\mathrm{comm}}]
          -|\mathcal N_\pi(u)\cap\mathcal N_\pi(v)|\leq0 .
 \label{eq:merge-pair-gain}
\end{equation}
These properties extend to batch merges.

A class pair is considered only if some pair of its instances communicates
directly or shares a neighbor, giving a positive pair-count reduction
in~\eqref{eq:merge-pair-gain}. This condition applies to the class pair,
so an individual merge may leave \(P\) unchanged. We also check that
the proposed composition has a compatible process and satisfies the
necessary area bounds.

For each remaining recipe, we select the instances for one new die at a
time. We start with an eligible pair of dies and add instances one by
one until the group matches the recipe. We choose each merge to minimize
\(B_{\mathrm{cut}}\), breaking ties by the largest reduction in \(P\).
If a tie remains, we favor the choice that leaves the most unused dies
with at least one of their best eligible partners. Along this sequence,
we evaluate the partition at each prescribed batch size \(m\).

After each merge, we reselect a process for each reuse class from the
technologies allowed by all its constituent blocks. All instances of the
class share this process. Selection prioritizes necessary physical
constraints and then minimizes the class's contribution to \(D\),
accounting for local-NoC area and design reuse. The resulting process
assignment is fixed for the downstream stages.

\subsection{Adaptive Pareto Exploration}
\label{sec:partition-search}

Algorithm~\ref{alg:partition} keeps the evaluated, pending, and expanded
states in \(\mathcal E,\mathcal U,\mathcal X\), respectively.
\(\operatorname{Admit}\) screens evaluated merge candidates before
\(\operatorname{Select}\) chooses representatives from them.
A nonnegative vector \(\boldsymbol v(s)\) records violations of necessary
conditions on the number of direct connections per chiplet, communication
reachability within the hop bounds, and available die-edge length for
interfaces. The first two conditions follow from the topology
restrictions in Section~\ref{sec:topology-domain}.
If \(D(s)\) exceeds the partition-stage cost ceiling in capacity mode,
\(e_D(s)\) equals the excess. Otherwise \(e_D(s)=0\).
A candidate state \(t\) generated from
state \(s\) is admitted for further search only if
\begin{equation}
 \boldsymbol v(t)\leq\boldsymbol v(s),\qquad e_D(t)\leq e_D(s).
 \label{eq:partition-admission}
\end{equation}
This allows successive merges to repair constraint violations without
worsening any individual violation or increasing the cost excess.

\(\operatorname{Select}\) uses one rule to choose how many pending states
to expand and how many admitted successors to queue per state.
On its input pool's Pareto frontier \(F_{\mathrm{pool}}\), it first selects
the candidates minimizing \(D\), \(B_{\mathrm{cut}}\), and \(P\).

For relative coverage, we subtract \(D_{\mathrm{fix}}\), the sum of
\(N_{\mathrm{IP}}/Q\) and the terms in~\eqref{eq:stage1-die-ledger}
for fixed input dies with prescribed processes, such as HBM stacks.
Let \(L(s,t)\) be the largest relative increase across these scores
when \(t\) represents \(s\).
\begin{equation}
 \begin{aligned}
 \tilde{\boldsymbol q}(s)&=(D(s)-D_{\mathrm{fix}},B_{\mathrm{cut}}(\pi),P(\pi)),\\
 L(s,t)&=\max_\ell\left[\frac{\tilde q_\ell(t)}{\tilde q_\ell(s)}-1\right]_+,\\
 \rho_j&=\max_{s\in F_{\mathrm{pool}}}\min_{t\in\mathcal S_j}L(s,t).
 \end{aligned}
 \label{eq:partition-relative-coverage}
\end{equation}
Here \([z]_+=\max(0,z)\). If \(\tilde q_\ell(s)=0\), the component loss is zero when
\(\tilde q_\ell(t)=0\) and infinite otherwise. Given the first \(j\)
representatives \(\mathcal S_j\), we add the state with the worst
remaining loss \(\rho_j\).

With \(j_0\) initial representatives and \(0<\rho_{j_0}<\infty\), we select
the shortest prefix maximizing coverage improvement per representative:
\begin{equation}
 \eta_j=\frac{1-\rho_j/\rho_{j_0}}{j},\qquad j\geq j_0 .
 \label{eq:partition-prefix-efficiency}
\end{equation}
If \(\rho_{j_0}=0\), retain the initial set; if
\(\rho_{j_0}=\infty\), retain the shortest prefix with finite loss.

In Algorithm~\ref{alg:partition}, \(\operatorname{Evaluate}\) assigns
processes to candidate partitions and computes
\(\boldsymbol q(s),\boldsymbol v(s),e_D(s)\).
Duplicate and detected symmetry-equivalent states are merged.

\begin{paperalgorithm}{alg:partition}{Adaptive Pareto Search for Chiplet Partitions}
\textbf{Input:} initial partition \(\pi_0\), traffic domain \(\Omega\),
constraints, technologies, \(Q\), optional partition-stage cost ceiling,
stopping parameters \(W,n_{\min},\varepsilon\).\\
\textbf{Output:} current \((D,B_{\mathrm{cut}},P)\) frontier \(F\)
of evaluated states satisfying \(\operatorname{Pass}\).\\
\textbf{Definitions:} \(\operatorname{Select}\):
\eqref{eq:partition-relative-coverage}--\eqref{eq:partition-prefix-efficiency};
\(\operatorname{Admit}\): \eqref{eq:partition-admission};
\(\operatorname{Pass}\): \(\boldsymbol v(s)=\boldsymbol0\) and \(e_D(s)=0\);
\(\operatorname{Converged}\): \eqref{eq:partition-frontier-gain} at checkpoints.
\par\smallskip
\begin{tabularx}{\linewidth}{@{}rY@{}}
1 & \(\mathcal E\gets\operatorname{Evaluate}(\{\pi_0\});\
     \mathcal U\gets\mathcal E;\ \mathcal X\gets\emptyset\)\\
2 & \textbf{while} \(\mathcal U\ne\emptyset\) \textbf{do}\\
3 & \quad \(\mathcal T\gets\operatorname{Select}(\mathcal U);\
     \mathcal U\gets\mathcal U\setminus\mathcal T\)\\
4 & \quad \textbf{for} \(s\in\mathcal T\) \textbf{do}\\
5 & \qquad \(\mathcal C\gets
      \operatorname{Evaluate}(\operatorname{BatchMerge}(s))\)\\
6 & \qquad \(\mathcal E\gets\mathcal E\cup\mathcal C\)\\
7 & \qquad \(\mathcal C\gets\operatorname{Admit}(\mathcal C,s)\)\\
8 & \qquad \(\mathcal U\gets\mathcal U\cup
      (\operatorname{Select}(\mathcal C)\setminus(\mathcal X\cup\mathcal T))\)\\
9 & \quad \textbf{end for}\\
10 & \quad \(\mathcal X\gets\mathcal X\cup\mathcal T\)\\
11 & \quad \(F\gets\operatorname{Pareto}_{\boldsymbol q}
      (\{s\in\mathcal E:\operatorname{Pass}(s)\})\)\\
12 & \quad \textbf{if} \(\operatorname{Converged}(F,|\mathcal X|)\)
      \textbf{then break}\\
13 & \textbf{end while}\\
14 & \textbf{return} \(F\)
\end{tabularx}
\end{paperalgorithm}

All evaluated states passing the partition checks and cost ceiling can
contribute to the output frontier, whether or not they are selected for
expansion. Representative selection controls which descendants are explored.

In Algorithm~\ref{alg:partition}, \(\operatorname{Converged}\) checks
progress at complete-batch boundaries once at least \(W\) additional states
have been expanded since the previous checkpoint. At checkpoint \(i\), let
\(F_i\) be the output frontier and \(n_i=|\mathcal X|\) the expanded-state
count. We normalize both \(F_{i-1}\) and \(F_i\) using the coordinate
ranges observed on output frontiers through checkpoint \(i\).
Let \(\operatorname{HV}_i(F)\) denote the hypervolume dominated by a
nonempty frontier \(F\) up to the reference point \((1.05,1.05,1.05)\)
in this normalized space.
The relative gain per \(W\) expansions is
\begin{equation}
 g_i=\frac{W}{\Delta n_i}
 \left[\frac{\operatorname{HV}_i(F_i)}{\operatorname{HV}_i(F_{i-1})}-1\right],
 \label{eq:partition-frontier-gain}
\end{equation}
where \(\Delta n_i=n_i-n_{i-1}\) counts newly expanded states.
\(\operatorname{Converged}\) stops the search after two consecutive
checkpoints with nonempty frontier pairs, \(n_i\geq n_{\min}\), and
\(g_i<\varepsilon\). In our experiments, we use \(W=100\), \(n_{\min}=200\),
and \(\varepsilon=10^{-3}\).
The final frontier is passed to topology synthesis and routing.

\section{Topology Synthesis and Routing}
\label{sec:stage2}

For a fixed partition and process assignment \((\pi,\boldsymbol\tau)\),
the topology synthesis and routing stage chooses how chiplets connect and carry the declared
traffic. It constructs a topology, determines routes and capacities, and
refines the topology to reduce interface resources while accounting for
die reuse and geometric risk. Algorithm~\ref{alg:topology} describes the
structural construction; routing and refinement follow below.

\subsection{Problem and Structural Search Domain}
\label{sec:topology-domain}

The chiplets form vertices \(V_\pi\); the search chooses undirected edges
\(E\) of a connected graph \(G=(V_\pi,E)\). For each \(G\), a fixed
shortest-path policy determines routing fractions \(\boldsymbol\alpha\).
The routes must respect the per-flow hop bounds, and the link capacities
\(\boldsymbol c\) must support every traffic vector in \(\Omega\).

\begin{figure}
  \centering
  \ResizePaperFigure{\columnwidth}{301.59924pt}{\begingroup\input{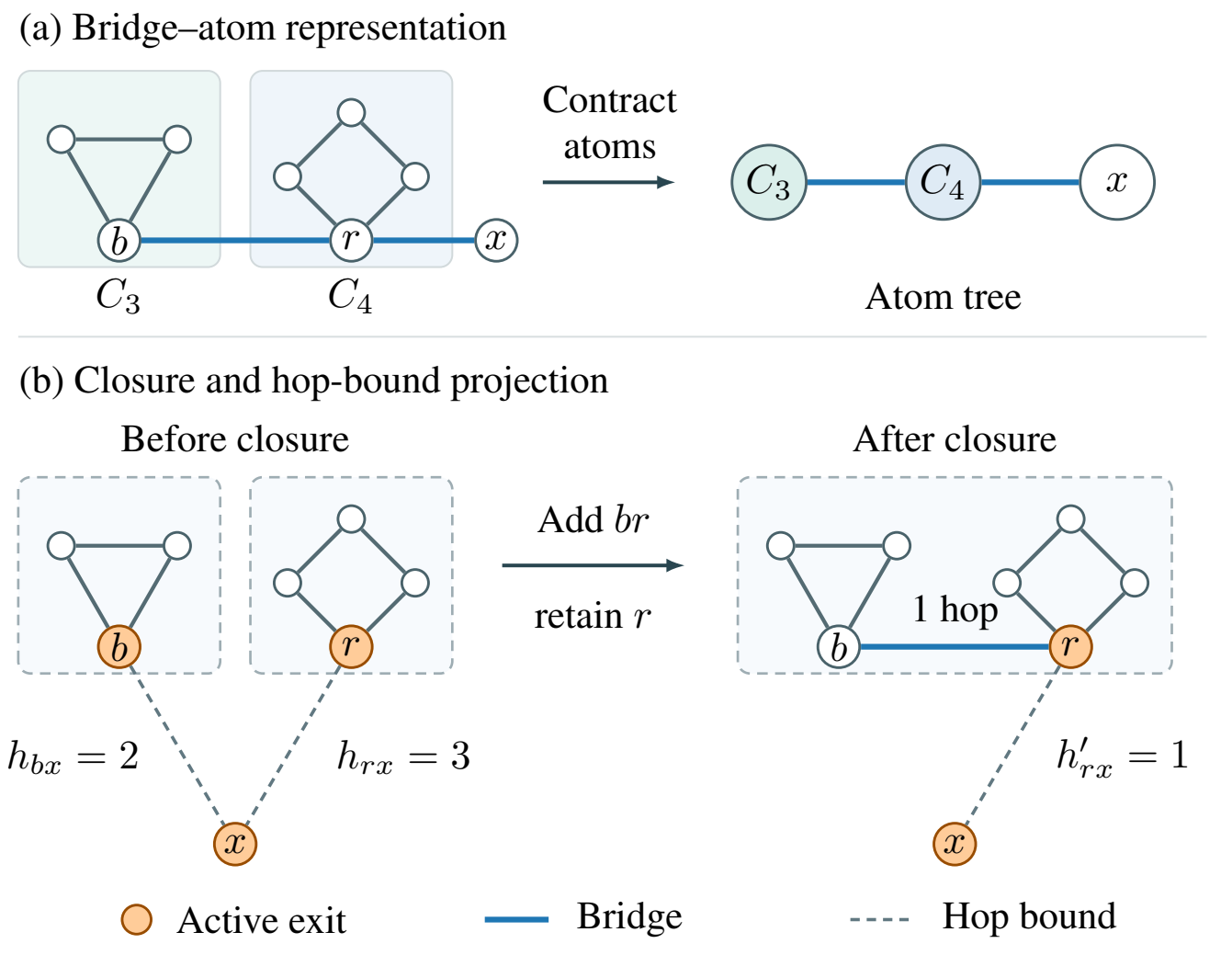}\tikzset{
  sTwoInner/.style={circle,draw=stageTwoInk!85,line width=.65pt,
    minimum size=.23cm,inner sep=0pt,fill=white},
  sTwoVertex/.style={sTwoInner,minimum size=.36cm,
    font=\PaperFigureFont,text=black},
  sTwoExit/.style={sTwoVertex,fill=orange!40,draw=orange!60!black},
  sTwoAtomArea/.style={draw=stageTwoGuide!35,line width=.45pt,
    rounded corners=3pt,fill=stageTwoPale},
  sTwoClosureArea/.style={draw=stageTwoGuide!75,line width=.6pt,
    dash pattern=on 2.6pt off 1.8pt,rounded corners=3pt,
    fill=stageTwoBlue!5},
  sTwoHop/.style={draw=stageTwoMuted,dash pattern=on 2.3pt off 1.7pt,
    line width=.8pt},
  sTwoSmall/.style={font=\PaperFigureFont,text=black,
    align=center,inner sep=1pt},
  sTwoHeading/.style={font=\PaperFigureFont,
    text=black,anchor=west,inner sep=0pt},
  sTwoBridge/.style={draw=stageTwoRoute,line width=1.4pt}
}\newcommand{\sTwoDrawAtoms}[3]{\foreach \name/\xx/\yy in {a/.55/.87,c/1.55/.87,d/2.5/.55,e/3.05/1.1,f/3.6/.55}
    \node[sTwoInner] (#1-\name) at (\xx,\yy) {};
  \node[#2] (#1-b) at (1.05,0) {$b$};
  \node[#3] (#1-r) at (3.05,0) {$r$};
  \draw[sTwoEdge] (#1-a)--(#1-b)--(#1-c)--(#1-a)
    (#1-r)--(#1-d)--(#1-e)--(#1-f)--(#1-r);
}\begin{tikzpicture}[sTwoScene,font=\PaperFigureFont]
  \path[use as bounding box] (0,.22) rectangle (10.6,8.42);
  \node[sTwoHeading] at (.18,8.16) {(a) Bridge--atom representation};

  \begin{scope}[shift={(0,6.35)}]
    \path[sTwoAtomArea,fill=stageTwoGreen!10] (.18,-.23) rectangle (1.92,1.46);
    \path[sTwoAtomArea,fill=stageTwoBlue!10] (2.18,-.23) rectangle (3.92,1.46);
    \sTwoDrawAtoms{top}{sTwoVertex}{sTwoVertex}
    \node[sTwoVertex] (top-x) at (4.3,0) {$x$};
    \draw[sTwoBridge] (top-b)--(top-r)--(top-x);
    \node[sTwoSmall] at (1.05,-.48) {$C_3$};
    \node[sTwoSmall] at (3.05,-.48) {$C_4$};
  \end{scope}
  \draw[sTwoFlow] (4.7,6.85)--(5.85,6.85);
  \node[sTwoSmall] at (5.28,7.37) {Contract\\atoms};
  \node[sTwoVertex,minimum size=.64cm,fill=stageTwoGreen!20] (atom-t) at (6.65,6.85) {$C_3$};
  \node[sTwoVertex,minimum size=.64cm,fill=stageTwoBlue!20] (atom-q) at (8.15,6.85) {$C_4$};
  \node[sTwoVertex,minimum size=.64cm] (atom-x) at (9.65,6.85) {$x$};
  \draw[sTwoBridge] (atom-t)--(atom-q)--(atom-x);
  \node[sTwoSmall] at (8.15,5.87) {Atom tree};

  \draw[draw=stageTwoGuide!30,line width=.45pt] (.18,5.52)--(10.42,5.52);
  \node[sTwoHeading] at (.18,5.12) {(b) Closure and hop-bound projection};
  \node[sTwoSmall] at (2.05,4.65) {Before closure};
  \node[sTwoSmall] at (8.25,4.65) {After closure};

  \begin{scope}[shift={(0,2.85)}]
    \path[sTwoClosureArea] (.18,-.23) rectangle (1.92,1.46);
    \path[sTwoClosureArea] (2.18,-.23) rectangle (3.92,1.46);
    \sTwoDrawAtoms{before}{sTwoExit}{sTwoExit}
  \end{scope}
  \node[sTwoExit] (before-x) at (2.05,1.15) {$x$};
  \draw[sTwoHop] (before-b)--(before-x) (before-r)--(before-x);
  \node[sTwoSmall] at (.65,1.9) {$h_{bx}=2$};
  \node[sTwoSmall] at (3.5,1.9) {$h_{rx}=3$};

  \draw[sTwoFlow] (4.35,3.55)--(5.93,3.55);
  \node[sTwoSmall] at (5.14,3.96) {Add $br$};
  \node[sTwoSmall] at (5.14,3.13) {retain $r$};

  \begin{scope}[shift={(6.2,2.85)}]
    \path[sTwoClosureArea] (.18,-.23) rectangle (3.92,1.46);
    \sTwoDrawAtoms{after}{sTwoVertex}{sTwoExit}
    \draw[sTwoBridge] (after-b)--(after-r);
    \node[sTwoSmall,text=black] at (2.05,.3) {1 hop};
  \end{scope}
  \node[sTwoExit] (after-x) at (8.25,1.15) {$x$};
  \draw[sTwoHop] (after-r)--(after-x);
  \node[sTwoSmall] at (9.7,1.9) {$h'_{rx}=1$};

  \begin{scope}[shift={(0,.5)}]
    \node[sTwoExit,minimum size=.25cm] at (1.2,0) {};
    \node[sTwoSmall,anchor=west] at (1.5,0) {Active exit};
    \draw[sTwoBridge] (4.2,0)--(4.75,0);
    \node[sTwoSmall,anchor=west] at (4.95,0) {Bridge};
    \draw[sTwoHop] (7.35,0)--(7.9,0);
    \node[sTwoSmall,anchor=west] at (8.1,0) {Hop bound};
  \end{scope}
\end{tikzpicture}\endgroup}
  \caption{(a) Contracting the \(C_3\) and \(C_4\) atoms yields a tree.
  (b) After bridge \(br\) joins the regions, the retained exit \(r\)
  must reach \(x\) in one hop.}
  \label{fig:stage2-atom-tree}
\end{figure}

A tree offers simple global path composition but only one path between
each pair. To control search complexity while retaining local connectivity
and routing flexibility, we deliberately restrict the topology search
space to a \emph{bridge--atom tree}: removing bridges leaves vertex-disjoint
atoms whose contraction forms a tree. We keep inter-atom connections
acyclic to reduce coupling between regions during downstream
floorplanning. The search uses singletons,
triangles (\(C_3\)), and chordless
four-cycles (\(C_4\))
(Fig.~\ref{fig:stage2-atom-tree}(a)). Triangles retain three-way direct
adjacency, while four-cycles provide two equal-length routes between
opposite vertices. This structure lets hop distances and routes compose
along the bridge tree. We also bound each chiplet's number of direct connections,
\(\deg_G(u)=|N_G(u)|\), where \(N_G(u)\) denotes its neighbors in~\(G\).

\subsection{Region Closure and Conflict Repair}
\label{sec:topology-closure}

Algorithm~\ref{alg:topology} constructs the topology by progressively joining
regions, starting with one chiplet per region. Each region keeps its
internal connections fixed and retains one chiplet, called its active
exit, for further connections.
A closure joins a set \(\mathcal W\) of active exits by a bridge or cycle, retains
\(r\in\mathcal W\), and seals the others; only \(r\) can acquire further external
connections for the combined region. Let \(h_{ux}\) be the current hop
bound between active exits, initialized from the tightest applicable
\(H_f\) and mandatory adjacency, or \(+\infty\) if unconstrained.
Let \(\delta(u,v)\) be the shortest-path hop distance and
\(\delta_{\mathrm{int}}(u,r)\) its value within the combined region.
Every path from \(u\in\mathcal W\) to an external exit \(x\) passes through \(r\),
so the hop distance and projected bound satisfy
\begin{equation}
 \begin{aligned}
 \delta(u,x)&=\delta_{\mathrm{int}}(u,r)+\delta(r,x),\\
 h'_{rx}&=\min_{u\in\mathcal W}\{h_{ux}-\delta_{\mathrm{int}}(u,r)\}.
 \end{aligned}
 \label{eq:closure-distance}
\end{equation}
The internal hop distances of a proposed closure must satisfy the current
bounds. We reject the closure if \(h'_{rx}<1\) for any external exit \(x\),
since reaching that exit requires at least one hop.
In Fig.~\ref{fig:stage2-atom-tree}(b), the new
bridge consumes one of the two hops allowed from \(b\) to \(x\), so the
retained exit \(r\) must connect directly to \(x\). Exit choices therefore
change the remaining hop bound. \(\operatorname{TryClosures}\) prefers closures
with smaller maximum tightening induced by communication demands. Secondary
preferences avoid sealing vertices with many remaining connection
opportunities. The first closure that passes propagation supplies the
neighbor facts \(\Delta\).

Sealing chiplet \(u\) records its complete physical neighbor set
\(N_G(u)\) as a \emph{neighbor fact}, fixing both its connections and
the exclusion of further neighbors. The accepted facts \(\mathcal A\)
determine state \(S\), including the regions, hop bounds, and edges \(E(S)\).
\(\operatorname{Propagate}\) rebuilds this state and checks
mandatory adjacency, connection caps, interface compatibility,
and atom structure. Using the projected bounds in
\eqref{eq:closure-distance}, it also rejects a partial topology when even
the remaining allowed connections cannot meet the hop bounds.

To repair a conflict, the search identifies a responsible combination
\(\Gamma\) of neighbor facts and adds it to the set of forbidden
combinations \(\mathcal N\) to prevent its recurrence. It withdraws the
accepted facts in \(\Gamma\) while preserving the others, then rebuilds
the regions and hop bounds to explore alternative closures.

\begin{paperalgorithm}{alg:topology}{Conflict-Directed Topology Construction}
\textbf{Input:} fixed \((\pi,\boldsymbol\tau)\), structural constraints;
search budget~\(W_c\).\\
\textbf{Output:} a structural candidate \(G\), or \textsc{unresolved}.\\
\textbf{Definitions:} \(\bot\): failure.\par\smallskip
\begin{tabularx}{\linewidth}{@{}rY@{}}
1 & \(\mathcal A\gets\emptyset;\ \mathcal N\gets\emptyset\)\\
2 & \textbf{while} \(\operatorname{WithinBudget}(W_c)\) \textbf{do}\\
3 & \quad \((S,\Gamma)\gets\operatorname{Propagate}(\mathcal A,\mathcal N)\)\\
4 & \quad \textbf{if} \(S\ne\bot\) \textbf{then}\\
5 & \qquad \textbf{if} \(\operatorname{Complete}(S)\) \textbf{then return} \(G=(V_\pi,E(S))\)\\
6 & \qquad \((\Delta,\Gamma)\gets\operatorname{TryClosures}(S,\mathcal A,\mathcal N)\)\\
7 & \qquad \textbf{if} \(\Delta\ne\bot\) \textbf{then} \(\mathcal A\gets\mathcal A\cup\Delta\); \textbf{continue}\\
8 & \quad \textbf{end if}\\
9 & \quad \textbf{if} \(\Gamma=\emptyset\) \textbf{then return} \textsc{unresolved}\\
10 & \quad \(\mathcal N\gets\mathcal N\cup\{\Gamma\};\
       \mathcal A\gets\mathcal A\setminus\Gamma\)\\
11 & \textbf{end while}\\
12 & \textbf{return} \textsc{unresolved}
\end{tabularx}
\end{paperalgorithm}

\subsection{Routing and Capacity Evaluation}

The returned structural candidate still requires routing and capacity
checks. For each candidate \(G\), traffic is split uniformly over its
interface-compatible shortest paths. Bridges, triangles, and adjacent \(C_4\) vertices have
unique local shortest paths; opposite \(C_4\) vertices split equally
over their admissible shortest arms. For each source--destination pair,
we compute link fractions on the directed acyclic graph formed by its
shortest admissible paths, without enumerating complete paths.

For each link \(e\), \(\alpha_e^{f,+}\) and \(\alpha_e^{f,-}\) are the
fixed fractions of flow \(f\) carried in the two directions.
The vectors \(\boldsymbol\alpha_e^\pm\) contain these fractions for all
flows. Over \(\Omega\), \(F_e\) and \(R_e\) are the peak bandwidths in
the two directions, and \(J_e\) is the peak total bandwidth across both
directions.
By~\eqref{eq:traffic-support},
\begin{align}
 F_e&=\sigma_\Omega(\boldsymbol\alpha_e^+),&
 R_e&=\sigma_\Omega(\boldsymbol\alpha_e^-),\nonumber\\
 J_e&=\sigma_\Omega(\boldsymbol\alpha_e^++\boldsymbol\alpha_e^-).
 \label{eq:edge-supports}
\end{align}
The interface model fixes each edge's transmit/receive organization;
package realization selects compatible link options.
A runtime-shared bidirectional interface needs \(c_e=J_e\); equal independent
transmit/receive resources need total capacity \mbox{\(c_e=2\max(F_e,R_e)\)}.
The total requirement \(B_{\mathrm{req}}(G)=\sum_{e\in E}c_e\) counts each
physical edge once.

\subsection{Topology Refinement}

Chiplets in the same reuse class \(g\in\mathcal G_\pi\) share one die
design, including its interface resources. A topology that concentrates
capacity requirements on a few instances can therefore increase the
interface resources replicated across the class.
Let \(B_u=\sum_{e\ni u}c_e\) be the sum of the required capacities of all
links connected to chiplet \(u\). Replicating the class peak
\(\bar B_g=\max_{u\in g}B_u\) in
every copy gives a bandwidth proxy for common interface provisioning,
\begin{equation}
 B_{\mathrm{reuse}}(G)=\sum_{g\in\mathcal G_\pi}n_g\bar B_g.
 \label{eq:stage2-reuse-capacity}
\end{equation}
Cost mode compares \((B_{\mathrm{reuse}},B_{\mathrm{req}})\)
lexicographically. Capacity mode reverses the order.

\begin{figure}
  \centering
  \ResizePaperFigure{\columnwidth}{297.33107pt}{\begingroup\input{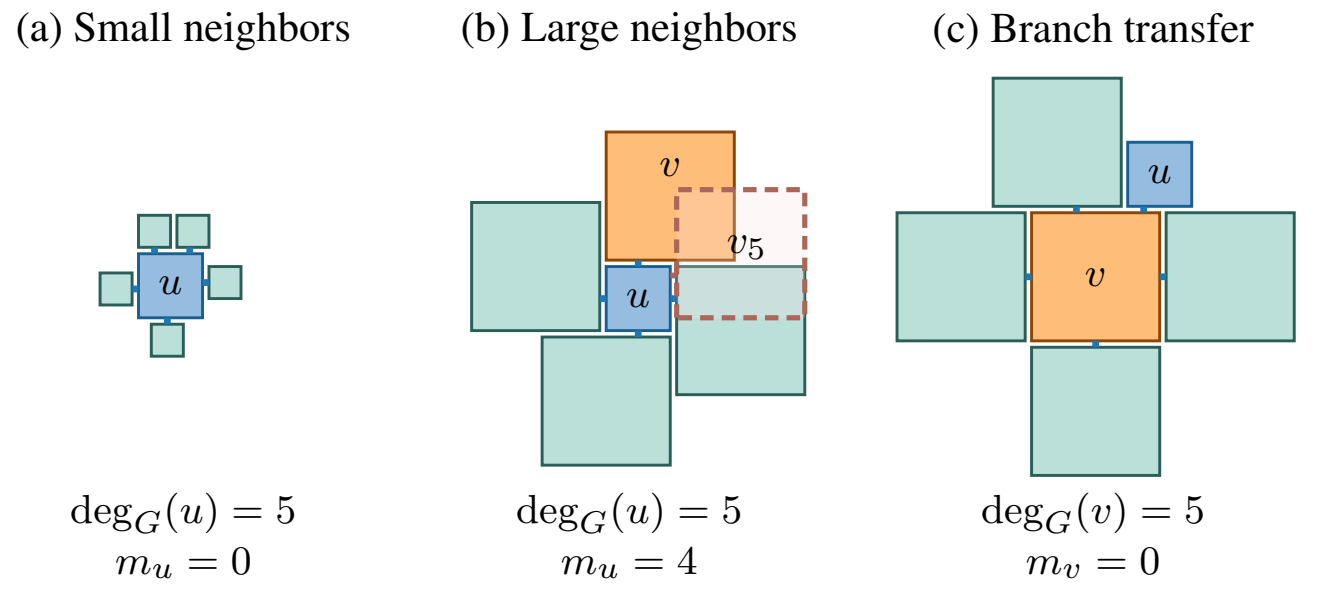}\tikzset{
  sTwoDie/.style={draw=stageTwoGreen!55!black,line width=.65pt,fill=stageTwoGreen!38},
  sTwoUnitDie/.style={sTwoDie,draw=stageTwoRoute!80!black,fill=stageTwoBlue!65},
  sTwoBranchDie/.style={sTwoDie,draw=orange!55!black,fill=orange!52},
  sTwoPhysicalLink/.style={draw=stageTwoRoute,line width=1.5pt},
  sTwoDieLabel/.style={font=\PaperFigureFont,text=black,
    inner sep=0pt},
  sTwoGeometryTitle/.style={font=\PaperFigureFont,
    text=black,align=center},
  sTwoGeometryNote/.style={font=\PaperFigureFont,
    text=black,align=center}
}\begin{tikzpicture}[x=1cm,y=1cm]
  \path[use as bounding box] (0,0) rectangle (10.45,4.6);
  \node[sTwoGeometryTitle] at (1.5,4.36) {(a) Small neighbors};
  \node[sTwoGeometryTitle] at (4.98,4.36) {(b) Large neighbors};
  \node[sTwoGeometryTitle] at (8.6,4.36) {(c) Branch transfer};

  \begin{scope}[shift={(1.15,2.12)},scale=.5]
    \path[sTwoUnitDie] (0,0) rectangle (1,1);
    \node[sTwoDieLabel] at (.5,.5) {$u$};
    \foreach \xx/\yy in {0/1.1,.6/1.1,1.1/.3,.2/-.6,-.6/.2}
      \path[sTwoDie] (\xx,\yy) rectangle ++(.5,.5);
    \draw[sTwoPhysicalLink] (.25,1)--(.25,1.1) (.8,1)--(.8,1.1)
      (1,.55)--(1.1,.55) (.45,0)--(.45,-.1) (0,.45)--(-.1,.45);
  \end{scope}
  \node[sTwoGeometryNote] at (1.5,.42) {$\deg_G(u)=5$\\$m_u=0$};

  \begin{scope}[shift={(4.8,2.02)},scale=.5]
    \foreach \xx/\yy in {1.1/-1,-1/-2.1,-2.1/0}
      \path[sTwoDie] (\xx,\yy) rectangle ++(2,2);
    \path[sTwoBranchDie] (0,1.1) rectangle ++(2,2);
    \node[sTwoDieLabel] at (1,2.55) {$v$};
    \path[sTwoUnitDie] (0,0) rectangle (1,1);
    \node[sTwoDieLabel] at (.5,.5) {$u$};
    \draw[sTwoPhysicalLink] (.5,1)--(.5,1.1) (1,.5)--(1.1,.5)
      (.5,0)--(.5,-.1) (0,.5)--(-.1,.5);
    \path[draw=stageTwoReject,fill=stageTwoPink!20,fill opacity=.25,
      line width=1.15pt,dash pattern=on 3pt off 1.8pt]
      (1.1,.2) rectangle ++(2,2);
    \draw[draw=stageTwoReject,line width=1.3pt] (1,.86)--(1.1,.86);
    \node[sTwoDieLabel,text=black] at (2.2,1.3) {$v_5$};
  \end{scope}
  \node[sTwoGeometryNote] at (4.98,.42) {$\deg_G(u)=5$\\$m_u=4$};

  \begin{scope}[shift={(8.12,1.94)},scale=.5]
    \foreach \xx/\yy in {-.6/2.1,2.1/0,0/-2.1,-2.1/0}
      \path[sTwoDie] (\xx,\yy) rectangle ++(2,2);
    \path[sTwoBranchDie] (0,0) rectangle (2,2);
    \node[sTwoDieLabel] at (1,1) {$v$};
    \path[sTwoUnitDie] (1.5,2.1) rectangle ++(1,1);
    \node[sTwoDieLabel] at (2,2.6) {$u$};
    \draw[sTwoPhysicalLink] (.7,2)--(.7,2.1) (1.75,2)--(1.75,2.1)
      (2,1)--(2.1,1) (1,0)--(1,-.1) (0,1)--(-.1,1);
  \end{scope}
  \node[sTwoGeometryNote] at (8.6,.42) {$\deg_G(v)=5$\\$m_v=0$};
\end{tikzpicture}\endgroup}
  \caption{A 2.5D example of the geometric risk score \(m_u\).
  (a) Small neighbors fit around \(u\).
  (b) Large neighbors block an additional attachment despite sufficient
  total shoreline.
  (c) Transferring the branch from \(u\) to larger die \(v\) yields a feasible
  layout with the same dies.}
  \label{fig:stage2-geometry-risk}
\end{figure}

Several large neighbors can be difficult to place around a smaller
chiplet (Fig.~\ref{fig:stage2-geometry-risk}). We use a score \(m_u\) to
estimate this risk and guide topology refinement. Relative linear sizes
are approximated from the base die areas \(A_u\). With
\(q_{uv}=\max\{0,\sqrt{A_v/A_u}-1\}\), define
\begin{equation}
 m_u=\begin{cases}
 \displaystyle\sum_{v\in N_G(u)}q_{uv}
 -\max_{v\in N_G(u)}q_{uv}, & \deg_G(u)>2,\\
 0, & \text{otherwise}.
 \end{cases}
 \label{eq:branch-size-risk}
\end{equation}
The largest neighbor is excluded so that a single large attachment
carries no penalty. \(m_u\) targets branches with multiple larger
neighbors.

We use the initial \(m_u\) values to guide refinement.
If \(\sum_u m_u<3\), it improves the capacity pair while
keeping the sum below 3. Otherwise, it first minimizes
\((\max_u m_u,\sum_u m_u)\), using the capacity pair to break ties.

A bounded first-improvement search transfers branching connections,
exchanges two edges, or rewrites a local three- or four-vertex region
while keeping its external attachments fixed. Each trial is checked
against structural and communication constraints, with routing and
capacities recomputed under the same policy.

\section{Physical Package Realization}
\label{sec:stage3}

\begin{figure*}
  \centering
  \ResizePaperFigure{\textwidth}{680.87393pt}{\input{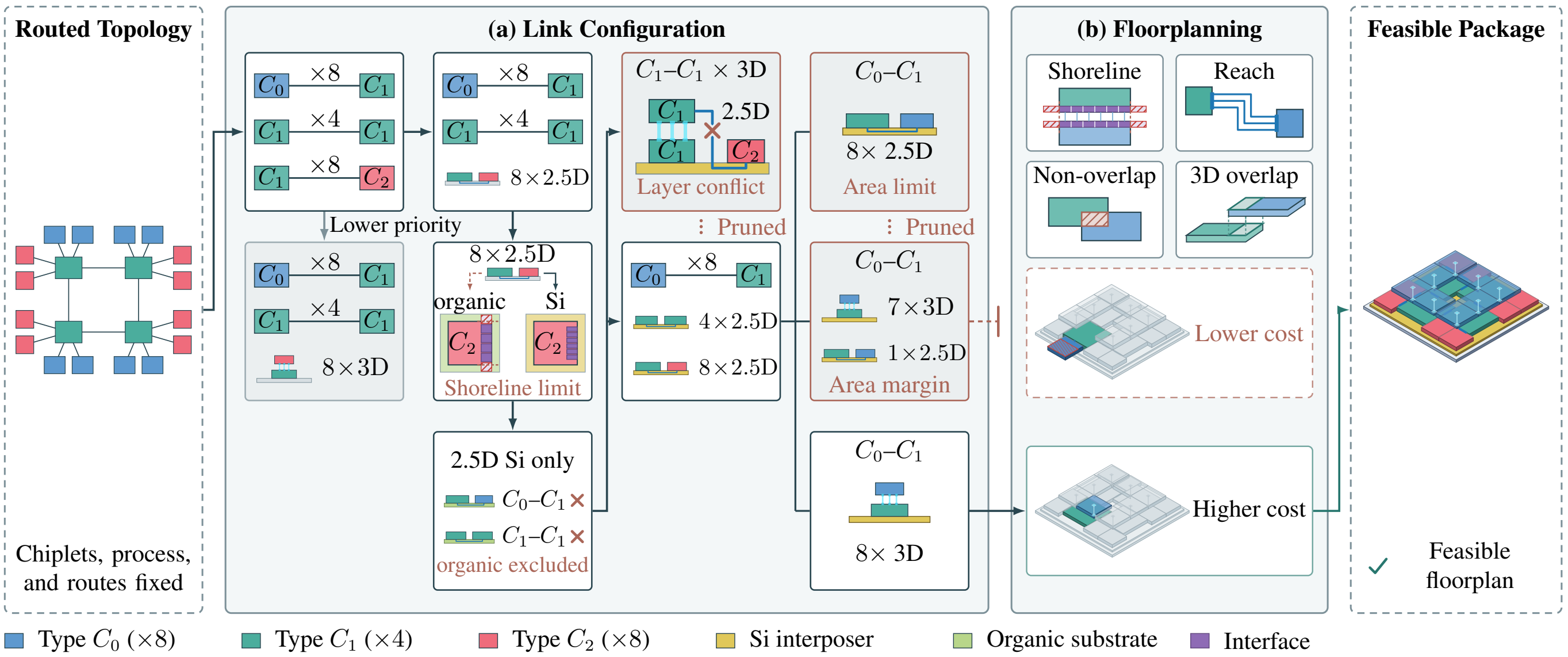}}
  \caption{Package realization for one routed topology.
  (a) We group link choices across carrier branches by the types of their endpoint chiplets.
  For \(C_1\!\text{--}\!C_2\), the lower bound on \(\widehat C\) favors 2.5D over 3D.
  Of the planar options, only Si meets the shoreline constraint.
  Layer conflicts exclude \(C_1\!\text{--}\!C_1\) stacking.
  For \(C_0\!\text{--}\!C_1\), the area limit rejects \(8\!\times\!\mathrm{2.5D}\) and the whitespace margin filters out
  \(7\!\times\!\mathrm{3D}+1\!\times\!\mathrm{2.5D}\), leaving \(8\!\times\!\mathrm{3D}\) for floorplanning.
  (b) Floorplanning checks shoreline use, planar reach (Manhattan distance
  between the midpoints of a link's two shoreline segments), non-overlap, and inter-layer overlap.}
  \label{fig:stage3-realization}
\end{figure*}

Package realization takes the partition, process assignments, topology,
and routing as fixed inputs. It selects a discrete link configuration
by cost, then solves its geometry.
Figure~\ref{fig:stage3-realization} illustrates the overall package
realization flow.

\subsection{Package Cost Model}
\label{sec:package-cost}

Package cost combines the die costs in~\eqref{eq:stage1-die-ledger}
with carrier and assembly costs. We include transit traffic when sizing
the local NoC. Die-design NRE covers integration, physical implementation,
masks, and test development. Recurring costs follow CATCH's manufacturing,
testing, and yield models~\cite{graening2026catch}.

We evaluate the package cost \(C(d,Q)\) on the final floorplan using
the die and carrier footprints, bonding areas, and CATCH's discrete
manufacturing calculations. For cost comparisons during search, we use a
continuous approximation \(\widehat{C}(d,Q)\). It retains integer
interface counts, relaxes wafer and reticle discretization, and smooths
stitching yield. Before floorplanning, it uses area lower bounds that
exclude layout whitespace. Both costs include amortized NRE.

CATCH's assembly hierarchy excludes a larger die spanning several smaller
dies. Motivated by experiments demonstrating a flat bonding surface across
dies and gap-fill material~\cite{lee2025reconstructed} and by proposed
staggered stacks~\cite{wang2026staggered}, we use an \emph{assembly graph}
in which an upper die can bond to multiple coplanar lower dies.
Each physical die--carrier or adjacent die--die bond constitutes one
assembly operation, aggregating the interface resources that share it.
For operation \(s\), let \(V_s\) be the participating groups' accumulated recurring cost;
\(a_s\) the assembly and
group-test expense; and \mbox{\(y_s\in(0,1]\)} the effective yield. The cost
increment is
\begin{equation}
 \Delta C_s=\frac{V_s+a_s}{y_s}-V_s
 =a_s+\underbrace{\left(y_s^{-1}-1\right)(V_s+a_s)}_{\text{yield-loss cost}}.
 \label{eq:assembly-cost-increment}
\end{equation}
Because the yield-loss term depends on the accumulated cost of the
groups being joined, we choose the assembly order by local dynamic
programming followed by insertion-based local search.
Additional alignment risk is modeled by a heuristic \(0.99\)
yield factor for each lower-die parent beyond the first.

A \emph{link option} pairs an interface implementation with a
compatible packaging technology. Different interfaces may use the
same technology, such as 2.5D packaging with a silicon interposer.
A \emph{link configuration} assigns an option to each edge and a
starting layer to each die. Planar links connect dies mounted directly
on a common carrier that supports the selected planar options. Vertical
links connect dies on adjacent layers.

Each interface instance for option \(t\) provides capacity \(b_t\).
Edge \(e\) with option \(t_e\) requires
\begin{equation}
 n_e=\left\lceil c_e/b_{t_e}\right\rceil
 \label{eq:interface-count}
\end{equation}
instances. The installed capacity is
\(B_{\mathrm{pkg}}(d)=\sum_{e\in E}n_e b_{t_e}\geq B_{\mathrm{req}}(G)\).
Copies of a reusable die share one shape and one interface template.
The template may need extra sites if copies require the same macro type
at different locations. Every reserved site contributes to each copy's
area and cost. Only instances assigned to links contribute to
\(B_{\mathrm{pkg}}\).

\subsection{Link Configuration Search}

For each routed topology, link-configuration branch-and-bound
(Fig.~\ref{fig:stage3-realization}(a)) first selects a root-carrier mode
(silicon interposer, organic substrate, or no planar links) to restrict the link
options, then refines these options and die starting layers.

When ranking link configurations, we ignore differences in installed
capacity due to rounding in~\eqref{eq:interface-count}, so capacity mode
also uses estimated cost \(\widehat{C}\). Before floorplanning, we
compute this estimate for a complete configuration using the maximum
count of each interface macro type across copies of the same die. We
rank and prune each partial assignment using a lower bound on \(\widehat{C}\) over
its compatible completions. These bounds combine area lower bounds
with optimistic link resources, costs, and yields. Assembly bounds
apply~\eqref{eq:assembly-cost-increment} within isolated connections or
small die groups, retaining local testing and yield losses while relaxing
global cost propagation.

Layer consistency and necessary area and interface-resource bounds
further restrict the choices. We exploit symmetries of the input topology
that preserve chiplet attributes and link requirements to reduce
equivalent assignments. The search prioritizes branches with lower cost
bounds and continues exploring unpruned alternatives within its budget
after finding a complete configuration. Of the complete configurations
found for each routed topology, only the one with the lowest estimated
cost proceeds to floorplanning. If the solver finds no feasible layout,
evaluation of that topology ends. To reduce this risk on 2.5D carriers,
we skip configurations if the lower bound on bottom-layer die area plus
a margin for layout whitespace (300~mm\(^2\) in our experiments)
exceeds the carrier area limit.

\subsection{Floorplanning}

For each selected configuration, floorplanning keeps the carrier, link
options, per-link counts \(n_e\), and die layers fixed. A mixed-integer
model selects die aspect ratios from a given discrete set and determines
planar coordinates and carrier dimensions.
Each die may be rotated by \(90^\circ\) to swap its width and height.
For each interface macro type, all copies of a reusable die must
reserve the same sites in the die's coordinate system. Each copy may
use a different subset for its links.

Each die has an integer height span \(h_v\), equal to one for ordinary dies.
A die \(v\) with fixed starting layer
\(\ell_v\) occupies \([\ell_v,\ell_v+h_v)\); its top must satisfy
\(\ell_v+h_v-1\leq L_{\max}\), where \(L_{\max}\) is the package layer limit.
Any two intersecting height intervals
require disjoint planar interiors, even when their starting layers differ.
The model also enforces planar shoreline and reach, and sufficient
overlap for adjacent-layer vertical connections
(Fig.~\ref{fig:stage3-realization}(b)). For a process subject to a reticle
limit, the final die rectangle, including the area of interfaces and
through-silicon vias, must
fit within the reticle width and height. The original carrier limits apply to
the actual footprint, including die spacing and layout whitespace.

Starting from the die floorplan, we place planar interfaces along die
edges and align vertical interfaces where stacked dies overlap. With
the link configuration fixed, we can reserve additional interface sites
in the shared template as needed to serve all copies. Once every link
has valid sites at both ends, we update \(\widehat{C}\) to include all
reserved resources. We refine the selected layout and evaluate \(C\)
using the resulting floorplan and assembly order.

\section{Evaluation}\label{sec:evaluation}

Across seven benchmarks, we compare NoPG with fixed
reference designs, a KaHyPar-based partitioning flow, and ChipletPart
on package cost and installed capacity. We test batch merging against
single-pair merging and assess the benefit of joint partition and
packaging search by restricting each choice in turn. A GA100 case study
shows how packaging options shape partitioning and process choices and
how production volume shifts the balance between recurring cost and
amortized NRE.

\subsection{Experimental Setup}\label{sec:methodology}

\subsubsection{Benchmarks and technology}
Table~\ref{tab:benchmarks} lists the blocks and partitions we model from
public descriptions of the Manual designs. Table~\ref{tab:search-settings}
gives the process options, physical limits, search settings, and hardware.
Where HBM is present, we retain Manual's generation and stack count and
model each stack as three package layers tall.

To construct \(\Omega\), we specify each client's cache and memory
destinations and its assumed access shares. For GB200, we split the reported
10~TB/s of total bandwidth between the reference dies~\cite{nvidia_blackwell_hc2024}
into two directional budgets of 5~TB/s each.
For MI300's Infinity Cache and Sapphire Rapids' last-level cache, one
package-wide budget treats clients symmetrically by summing their read
and write rates.
We set this budget to the largest value for which, under the routing assumed
for Manual, the worst-case load in each link direction does not exceed its
nominal capacity.
These budgets apply to the same logical flows under every partition.
In our six-stack GA100 model,
each streaming multiprocessor (SM) accesses all L2 slices uniformly. The two
memory controllers (MCs) serving each HBM2 stack share 311~GB/s for reads and writes,
derived from 1555~GB/s across the five active stacks of the
\mbox{40-GB} A100~\cite{nvidia2020ampere}.
HBM stacks connect directly to their MCs. Selected compute--cache and
cache--MC flows (e.g., CU--L2 and IC--MC in MI300) have hop bound
\(H_f=1\). All remaining flows have \(H_f=3\).
Our library models separate transmit and receive lanes for UCIe Standard
on organic substrates, UCIe Advanced on silicon interposers, and UCIe-3D
between stacked dies~\cite{ucie30}. For Sapphire Rapids, we use UCIe Advanced
links on a 2.5D silicon interposer as a proxy for the EMIB connections
between the four reference tiles. HBM reads and writes share the data pins
and bandwidth of a dedicated planar interface. Each stack's equivalent
data port lies at the midpoint of one edge and rotates with the stack.
\begin{table}[!t]
  \caption{Modeled block inventories and fixed Manual partitions.}
  \label{tab:benchmarks}
  \centering\footnotesize
  \setlength{\tabcolsep}{2.5pt}
  \begin{tabularx}{\columnwidth}{@{}>{\raggedright\arraybackslash}p{0.20\columnwidth}Y>{\raggedright\arraybackslash}p{0.25\columnwidth}@{}}
    \toprule
    Benchmark & Non-HBM blocks by library node & Total blocks and Manual partition \\
    \midrule
    Intel Sapphire Rapids\newline\cite{intel_spr_uncore,intel8490h}
      & \textit{7 nm:} 60 CPU cores, 60 LLC slices, 4 DSA, 4 IAA, 4 QAT, 4 DLB, 4 dual-channel DDR5 controllers, 8 PCIe~x16 interfaces, 4 UPI interfaces
      & 152 blocks\newline 4 XCC tiles \\
    \addlinespace[3pt]
    AMD EPYC~7452\newline\cite{amd2020epyc,lenovo2023rome}
      & \textit{7 nm:} 32 CPU cores, 32 four-MiB L3 slices.\newline\textit{16 nm:} 4 dual-channel DDR4 controllers, 8 PCIe~x16 interfaces
      & 76 blocks\newline 4~CCDs, \mbox{1 I/O die} \\
    \addlinespace[3pt]
    AMD MI300X\newline\cite{smith2024mi300x}
      & \textit{5 nm:} 320 physical CUs, 128 L2 channels.\newline\textit{6 nm:} 128 IC channels, 8 HBM controllers, 4 IF and 4 IF/PCIe interfaces
      & 600 blocks\newline 8~XCDs, \mbox{4 I/O dies}\newline 8~HBM3 stacks \\
    \addlinespace[3pt]
    AMD MI300A\newline\cite{amd2025cdna3}
      & \textit{5 nm:} 240 physical CUs, 96 L2 channels, 24 CPU cores, 24 L3 slices.\newline\textit{6 nm:} same blocks as MI300X
      & 536 blocks\newline 6~XCDs, 3~CCDs, \mbox{4 I/O dies}\newline 8~HBM3 stacks \\
    \addlinespace[3pt]
    NVIDIA GA100\newline\cite{nvidia2020ampere}
      & \textit{7 nm:} 128 SMs, 96 L2 slices, 12 HBM controllers (512-bit), 12 NVLink interfaces, 1 PCIe interface
      & 255 blocks\newline 1 GPU die\newline 6~HBM2 stacks \\
    \addlinespace[3pt]
    NVIDIA GH100\newline\cite{nvidia2023hopper}
      & \textit{5 nm:} 144 SMs, 96 L2 slices, 12 HBM controllers (512-bit), 18 NVLink interfaces, 1 PCIe interface
      & 277 blocks\newline 1 GPU die\newline 6~HBM3 stacks \\
    \addlinespace[3pt]
    NVIDIA GB200 GPU\newline\cite{nvidia_blackwell_hc2024,nvidia2025b200sm}
      & \textit{5 nm:} 160 SMs, 128 L2 slices, 8 HBM controllers, 18 NVLink interfaces, 2 PCIe interfaces
      & 324 blocks\newline 2 GPU dies\newline 8~HBM3e stacks \\
    \bottomrule
  \end{tabularx}
\end{table}

\begin{table}
  \caption{Experimental settings.}
  \label{tab:search-settings}
  \centering\footnotesize
  \setlength{\tabcolsep}{4pt}
  \begin{tabularx}{\columnwidth}{@{}>{\raggedright\arraybackslash}p{0.35\columnwidth}Y@{}}
    \toprule
    Parameter & Setting \\
    \midrule
    \multicolumn{2}{@{}l}{\textit{Benchmark inputs and physical limits}} \\
    Process-node library & \(\{3,5,6,7,12,16,28\}\)~nm \\
    Die reticle & \(26\times33\)~mm (858~mm\(^2\)) \\
    NoPG die aspect ratios & \(1{:}1, 26{:}33, 2{:}3, 1{:}2, 1{:}3\) (either orientation) \\
    2.5D carrier limits & \(120\times120\)~mm maximum for both carriers. Si area cap 2574~mm\(^2\) (GB200~GPU proxy \(3.5\times830=2905\)~mm\(^2\)~\cite{tsmc2024cowosl}). \\
    Maximum planar reach & Si interposer 2~mm. Organic substrate 25~mm for UCIe and 6~mm for HBM. \\
    \midrule
    \multicolumn{2}{@{}l}{\textit{NoPG settings and execution environment}} \\
    Max.\ neighbors per chiplet & 6 (7 for MI300A to match Manual) \\
    NoPG parallelism & 48 threads \\
    Machine & AMD Threadripper 3960X, 128~GB RAM \\
    Solver & Gurobi~13.0.1 \\
    \bottomrule
  \end{tabularx}
\end{table}

\begin{table*}[!tp]
  \caption{Package cost \(C\) (USD per package) and installed capacity \(B_{\mathrm{pkg}}\) (TB/s). The pure 2.5D comparison uses Manual's carrier under relaxed floorplanning constraints.}
  \label{tab:qor-comparison}
  \centering\footnotesize
  \setlength{\tabcolsep}{1.2pt}
  \renewcommand{\arraystretch}{1.08}
  \begin{tabular*}{\textwidth}{@{\extracolsep{\fill}}ll*{8}{r}@{\hskip .7em}*{6}{r}@{}}
    \toprule
    \multirow{3}{*}{Benchmark} & \multirow{3}{*}{Mode} & \multicolumn{8}{c}{\(Q=10^6\)} & \multicolumn{6}{c}{Pure 2.5D, \(Q\to\infty\)} \\
    \cmidrule(lr){3-10}\cmidrule(l){11-16}
     & & \multicolumn{2}{c}{Manual} & \multicolumn{3}{c}{\(\mathrm{NoPG}_{\mathrm{KaHyPar}}\)} & \multicolumn{3}{c}{NoPG} & \multicolumn{3}{c}{ChipletPart} & \multicolumn{3}{c}{NoPG-C} \\
    \cmidrule(lr){3-4}\cmidrule(lr){5-7}\cmidrule(lr){8-10}\cmidrule(lr){11-13}\cmidrule(l){14-16}
     & & \(C\) & \(B_{\mathrm{pkg}}\) & \(C\) & \(B_{\mathrm{pkg}}\) & Time (s) & \(C\) & \(B_{\mathrm{pkg}}\) & Time (s) & \(C\) & \(B_{\mathrm{pkg}}\) & Time (s) & \(C\) & \(B_{\mathrm{pkg}}\) & Time (s) \\
    \midrule
    \multirow{2}{*}{\shortstack[l]{Intel\\Sapphire Rapids}} & C & \multirow{2}{*}{986.24} & \multirow{2}{*}{10.240} & 833.47 & 36.208 & 131.9 & 629.71 & 34.864 & 282.6 & \multirow{2}{*}{634.32} & \multirow{2}{*}{30.208} & \multirow{2}{*}{2,568.5} & \multirow{2}{*}{569.94} & \multirow{2}{*}{29.696} & \multirow{2}{*}{1,537.1} \\
     & B &  &  & 934.31 & 27.328 & 129.4 & 978.83 & 5.920 & 48.2 &  &  &  &  &  &  \\
    \midrule
    \multirow{2}{*}{\shortstack[l]{AMD\\EPYC 7452}} & C & \multirow{2}{*}{291.55} & \multirow{2}{*}{0.512} & 193.41 & 8.528 & 79.6 & 170.51 & 3.040 & 77.4 & \multirow{2}{*}{141.39} & \multirow{2}{*}{1.024} & \multirow{2}{*}{1,009.2} & \multirow{2}{*}{138.23} & \multirow{2}{*}{1.024} & \multirow{2}{*}{13.3} \\
     & B &  &  & 288.31 & 2.240 & 19.9 & 278.74 & 0.480 & 3.7 &  &  &  &  &  &  \\
    \addlinespace[1pt]
    \multirow{2}{*}{\shortstack[l]{AMD\\MI300X}} & C & \multirow{2}{*}{3,609.22} & \multirow{2}{*}{62.413} & 3,746.56 & 67.469 & 957.5 & 3,467.95 & 86.413 & 492.2 & \multirow{2}{*}{4,205.36} & \multirow{2}{*}{40.141} & \multirow{2}{*}{4,468.0} & \multirow{2}{*}{3,317.18} & \multirow{2}{*}{77.517} & \multirow{2}{*}{780.5} \\
     & B &  &  & \multicolumn{2}{c}{\(\times\)} & 932.9 & 3,579.47 & 38.093 & 730.2 &  &  &  &  &  &  \\
    \addlinespace[1pt]
    \multirow{2}{*}{\shortstack[l]{AMD\\MI300A}} & C & \multirow{2}{*}{3,622.84} & \multirow{2}{*}{54.733} & 3,574.27 & 59.277 & 769.5 & 3,487.69 & 60.365 & 548.3 & \multirow{2}{*}{5,757.98} & \multirow{2}{*}{42.189} & \multirow{2}{*}{3,695.7} & \multirow{2}{*}{3,151.19} & \multirow{2}{*}{56.013} & \multirow{2}{*}{677.7} \\
     & B &  &  & \multicolumn{2}{c}{\(\times\)} & 754.4 & 3,617.85 & 45.613 & 928.6 &  &  &  &  &  &  \\
    \midrule
    \multirow{2}{*}{\shortstack[l]{NVIDIA\\GA100}} & C & \multirow{2}{*}{2,307.60} & \multirow{2}{*}{1.866} & 1,835.23 & 60.122 & 171.3 & 1,791.87 & 65.578 & 240.5 & \multicolumn{2}{c}{\multirow{2}{*}{\(\times\)}} & \multirow{2}{*}{1,923.3} & \multirow{2}{*}{1,720.13} & \multirow{2}{*}{34.122} & \multirow{2}{*}{200.1} \\
     & B &  &  & 2,160.35 & 1.866 & 127.9 & 2,160.35 & 1.866 & 111.2 &  &  &  &  &  &  \\
    \addlinespace[1pt]
    \multirow{2}{*}{\shortstack[l]{NVIDIA\\GH100}} & C & \multirow{2}{*}{3,161.90} & \multirow{2}{*}{4.020} & 2,073.76 & 89.796 & 211.9 & 1,998.97 & 79.924 & 333.2 & \multirow{2}{*}{2,096.74} & \multirow{2}{*}{44.980} & \multirow{2}{*}{2,103.2} & \multirow{2}{*}{1,933.93} & \multirow{2}{*}{56.244} & \multirow{2}{*}{1,455.8} \\
     & B &  &  & 3,103.37 & 4.020 & 158.2 & 3,103.37 & 4.020 & 102.7 &  &  &  &  &  &  \\
    \addlinespace[1pt]
    \multirow{2}{*}{\shortstack[l]{NVIDIA\\GB200 GPU}} & C & \multirow{2}{*}{4,743.83} & \multirow{2}{*}{18.240} & 4,655.91 & 83.904 & 403.1 & 3,119.52 & 105.280 & 676.7 & \multicolumn{2}{c}{\multirow{2}{*}{\(\times\)}} & \multirow{2}{*}{3,616.3} & \multirow{2}{*}{3,175.11} & \multirow{2}{*}{107.840} & \multirow{2}{*}{1,398.7} \\
     & B &  &  & 4,655.91 & 83.904 & 398.7 & 4,741.07 & 18.240 & 406.9 &  &  &  &  &  &  \\
    \midrule
    \multirow{2}{*}{Geo. mean$^\dagger$} & C & \multirow{2}{*}{1.000} & \multirow{2}{*}{1.000} & 0.840 & 5.828 &  & \textbf{0.731} & 5.344 &  & \multirow{2}{*}{1.000} & \multirow{2}{*}{1.000} & \multirow{2}{*}{1.000} & \multirow{2}{*}{\textbf{0.811}} & \multirow{2}{*}{1.258} & \multirow{2}{*}{0.177} \\
     & B &  &  & 0.967 & 2.218 &  & 0.979 & \textbf{0.832} &  &  &  &  &  &  &  \\
    \bottomrule
  \end{tabular*}
  \par\smallskip
  \begin{minipage}{\textwidth}\footnotesize
  \mbox{NoPG-C} targets cost. \mbox{NoPG-B} targets installed capacity, with final package cost capped at Manual's for the same \(Q\). \(\times\) No valid design reported.
  \par \(^{\dagger}\) Ratios to Manual use each method's valid cases. Pure 2.5D ratios to ChipletPart use cases both tools report as valid (5 benchmarks).
  \end{minipage}
\end{table*}

We calibrate block areas to reference dies using public die and cache
specifications and component models~\cite{zhang2024llmcompass}.
We model Sapphire Rapids' Intel~7 at 7~nm, GH100's TSMC~4N and GB200's
TSMC~4NP at 5~nm, and EPYC's 14-nm I/O die at 16~nm~\cite{amd2020epychpc}.
Logic and SRAM areas scale separately
with TSMC density and SRAM-cell data~\cite{tsmc_advanced_technologies,yeap2019n5,chen2015sram}.
Analog area stays fixed. From TMAC's linear NoC model~\cite{wang2024tmac}
and Theseus's 14-nm router data~\cite{zhu2025theseus}, we estimate
\(\kappa_{\mathrm{NoC}}(16\,\mathrm{nm})\approx
1.486\times10^{-3}\,\mathrm{mm}^2/(\mathrm{GB/s})\) and scale it with
logic area.
Our common evaluator uses CATCH's manufacturing, test, and yield
models~\cite{graening2026catch} with published design and wafer costs
and 3-nm estimates~\cite{palma2022design,feng2022chipletactuary}.
For~\eqref{eq:stage1-die-ledger}, we allocate half the non-mask design
cost rate to reusable IP development, charged once per block type at
its reference process, and the other half to die implementation.

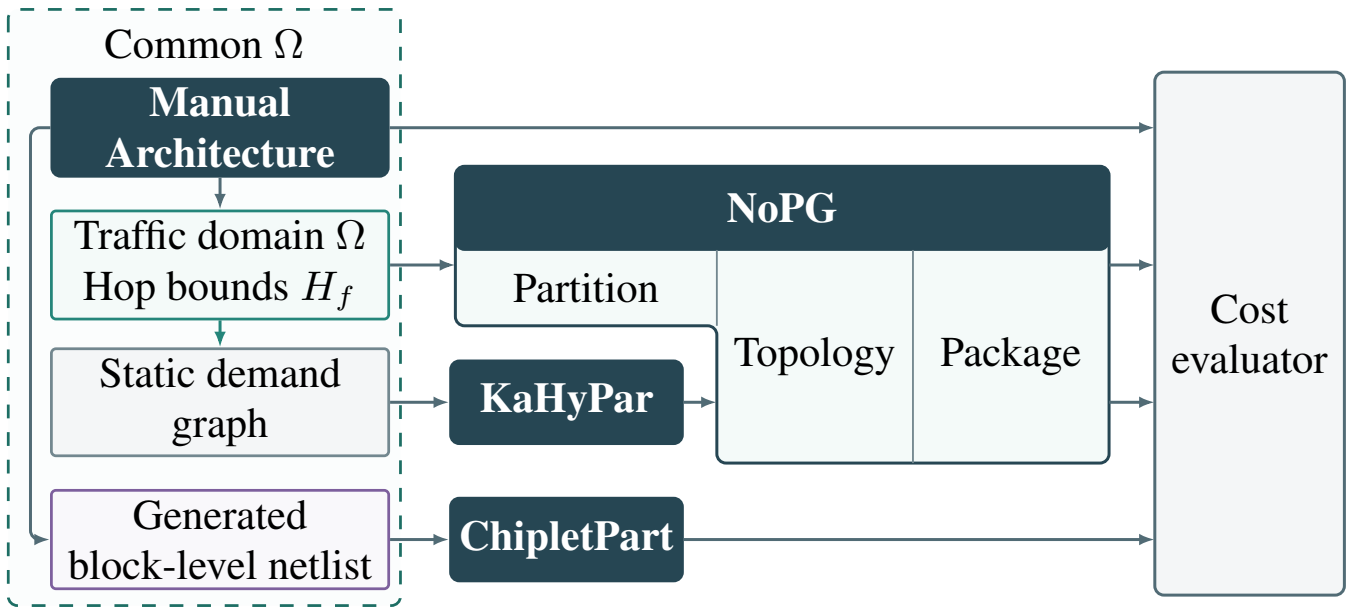
\begin{figure}[t]
  \centering
  \begingroup
  \definecolor{inputNavy}{HTML}{264653}
  \definecolor{inputTeal}{HTML}{2A9D8F}
  \definecolor{inputPurple}{HTML}{8B6AAE}
  \definecolor{inputGray}{HTML}{F4F6F7}
  \ResizePaperFigure{\columnwidth}{291.4pt}{\begin{tikzpicture}[
    x=1cm,y=1cm,font=\PaperFigureFont,text=black,
    input/.style={draw=inputNavy!65,fill=inputGray,rounded corners=1pt,
      minimum width=2.58cm,minimum height=.72cm,inner sep=2pt,
      align=center,line width=.6pt},
    method/.style={draw=inputNavy,fill=inputNavy,text=white,
      rounded corners=2pt,minimum width=1.78cm,minimum height=.64cm,
      font=\PaperFigureFont\bfseries,inner sep=2pt,align=center},
    arrow/.style={-{Latex[length=1.2mm]},draw=inputNavy!80,line width=.65pt}
  ]
    \path[draw=inputTeal!70!black,fill=inputTeal!2,dashed,
      rounded corners=2pt,line width=.6pt] (-.62,2.48) rectangle (2.38,-2.08);
    \node at (.88,2.22) {Common $\Omega$};

    \node[method,minimum width=2.58cm,minimum height=.72cm] (manual) at (1,1.575)
      {Manual\\Architecture};
    \node[input,draw=inputTeal!85!black,fill=inputTeal!5] (native) at (1,.525)
      {Traffic domain $\Omega$\\Hop bounds $H_f$};
    \node[input] (graph) at (1,-.525)
      {Static demand\\graph};
    \node[input,draw=inputPurple!90!black,fill=inputPurple!5] (netlist) at (1,-1.575)
      {Generated\\block-level netlist};

    \node[method] (kahypar) at (3.65,-.525) {KaHyPar};
    \node[method] (chipletpart) at (3.65,-1.575) {ChipletPart};
    \draw[arrow] (native.east) -- (2.80,.525);
    \draw[arrow] (graph) -- (kahypar);
    \draw[arrow] (netlist) -- (chipletpart);

    \draw[arrow] (manual.south) -- (native.north);
    \draw[arrow,draw=inputTeal!85!black] (native.south) -- (graph.north);
    \draw[arrow,rounded corners=2pt] (manual.west) -- (-.45,1.575) |- (netlist.west);

    \path[draw=inputNavy,fill=inputTeal!5,rounded corners=2pt,line width=.65pt]
      (2.80,1.28) -- (7.80,1.28) -- (7.80,-.99) -- (4.80,-.99)
      -- (4.80,.06) -- (2.80,.06) -- cycle;
    \node[method,minimum width=5.00cm] at (5.30,.96) {NoPG};
    \draw[draw=inputNavy!65,line width=.4pt] (4.80,.64) -- (4.80,.06);
    \draw[draw=inputNavy!65,line width=.4pt] (6.30,.64) -- (6.30,-.99);
    \node at (3.80,.35) {Partition};
    \node at (5.55,-.175) {Topology};
    \node at (7.05,-.175) {Package};
    \draw[arrow] (kahypar.east) -- (4.80,-.525);

    \path[draw=inputNavy!80,fill=inputGray,rounded corners=2pt,line width=.65pt]
      (8.15,2.00) rectangle (9.60,-2.00);
    \node[align=center] at (8.875,0)
      {Cost\\evaluator};
    \draw[arrow] (manual.east) -- (8.15,1.575);
    \draw[arrow] (7.80,.525) -- (8.15,.525);
    \draw[arrow] (7.80,-.525) -- (8.15,-.525);
    \draw[arrow] (chipletpart.east) -- (8.15,-1.575);
  \end{tikzpicture}}
  \endgroup
  \caption{Comparison flows with a common fixed-layout cost model.}
  \label{fig:evaluation-input-models}
\end{figure}

\subsubsection{Comparison flows}
\label{sec:external-compatibility}
We evaluate Manual as given and derive the other inputs in
Fig.~\ref{fig:evaluation-input-models} from its functional blocks, flow
endpoints, and traffic domain \(\Omega\). NoPG uses the domain directly.
To isolate the effect of partitioning, \(\mathrm{NoPG}_{\mathrm{KaHyPar}}\)
replaces Algorithm~\ref{alg:partition} with
KaHyPar~v1.3.7~\cite{gottesburen2020advanced}.
In its demand graph, we weight each communicating block pair by its peak
combined traffic in both directions over \(\Omega\). For balance, each
vertex is weighted by the minimum total area of its blocks across allowed
processes. We scan nonfixed partition counts from the reticle-area lower
bound to the vertex count using KaHyPar's direct \(k\)-way cut preset
(imbalance 0.5, seed 0). Both flows use NoPG for process assignment,
topology synthesis and routing, and package realization.

We use ChipletPart as our co-design baseline because its joint
search over chiplet partitions and process assignments produces
floorplans we can evaluate with the same cost model.
For~each die pair, ChipletPart sums the specified net bandwidths to size
its interfaces. Using block-pair peaks as net bandwidths can overstate capacity
when those peaks cannot coincide. We therefore generate a block-level netlist
of Manual's interconnect in which switches let flows share fixed connections.
We convert each reference die's peak load over \(\Omega\) to an equivalent
local-NoC area using~\eqref{eq:local-noc-area}, then divide it among the
die's switches in proportion to their peak loads. Each HBM stack remains a
separate die with its reference process and dimensions, so we exclude it
from ChipletPart's partition count.
We use ChipletPart's default genetic search settings and the comparison
script's default of 16 OpenMP threads. In pilot runs, a higher partition
cap gave no consistent savings, and a higher thread count yielded
costlier layouts or no valid layout.
For a fair comparison, both tools use Manual's 2.5D carrier and the
same planar interfaces.
ChipletPart supports only recurring-cost minimization, so we use
\mbox{NoPG-C} at \(Q\to\infty\), where amortized NRE per package vanishes.
Following ChipletPart's floorplanning model, we lift the carrier and
logic-die size limits and leave interface locations implicit.

\subsection{Comparison with Baselines}
\label{sec:results}

\begin{table}[tb]
  \centering\footnotesize
  \caption{Cost per package (USD) for three \mbox{NoPG-C} search
  settings at \(Q=10^6\).}
  \label{tab:packaging-choices}
  \setlength{\tabcolsep}{2pt}
  \begin{tabular*}{\columnwidth}{@{\extracolsep{\fill}}lrrr@{}}
    \toprule
    Benchmark & Fixed partition & Restricted packaging & Joint search \\
    \midrule
        Intel Sapphire Rapids & 866.01 & 782.92 & 629.71 \\
    AMD EPYC 7452 & 238.29 & 196.38 & 170.51 \\
    AMD MI300X & 3,872.06 & 3,750.41 & 3,467.95 \\
    AMD MI300A & 3,756.51 & 3,552.77 & 3,487.69 \\
    NVIDIA GA100 & 2,160.35 & 1,820.74 & 1,791.87 \\
    NVIDIA GH100 & 3,103.37 & 2,086.98 & 1,998.97 \\
    NVIDIA GB200 GPU & 4,741.07 & 4,581.66 & 3,119.52
\\
    \bottomrule
  \end{tabular*}
  \par\smallskip
  \begin{minipage}{\columnwidth}\footnotesize
    \emph{Fixed partition} searches the full packaging library with Manual's
    partition, while \emph{Restricted packaging} allows repartitioning
    within Manual's technologies (organic 2.5D for EPYC, silicon 2.5D
    and 3D for both MI300 models, and silicon 2.5D otherwise). All three
    settings reselect logic-die processes and synthesize a routed topology.
  \end{minipage}
\end{table}

\begin{figure*}[!t]
  \centering
  \includegraphics[width=\textwidth]{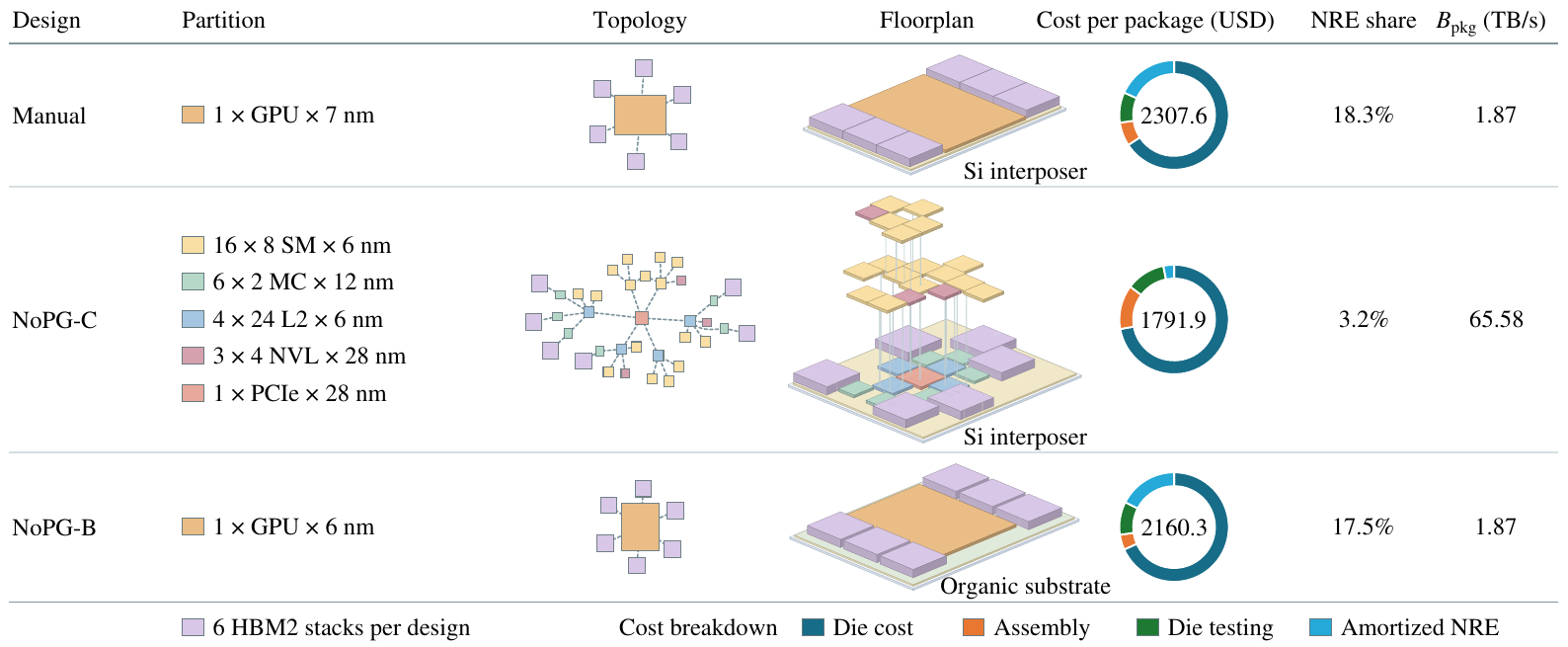}
  \caption{Chiplet partitions, topologies, and floorplans of the Manual, \mbox{NoPG-C},
  and \mbox{NoPG-B} designs for NVIDIA GA100 at \(Q=10^6\).
  Die cost covers yield-adjusted die fabrication before package interfaces are added.
  Assembly includes carrier fabrication, added die area for interconnects, bonding,
  group testing, and losses from failed assembly steps.}
  \label{fig:ga100-objectives}
\end{figure*}

\subsubsection{Manual reference}
At \(Q=10^6\), \mbox{NoPG-C} found the lowest reported package cost
on all seven benchmarks in 77.4--676.7~s per benchmark
(Table~\ref{tab:qor-comparison}). Lower amortized NRE accounts for 50.0\% of
the combined dollar savings over Manual. On GA100, \mbox{NoPG-C} splits
Manual's near-reticle-sized logic die into smaller, reusable dies,
lowering package cost by 22.3\%, or 51.8\% after subtracting the
standalone cost of the six fixed HBM2 stacks from both totals.
Under each Manual design's cost ceiling, \mbox{NoPG-B} found the lowest
reported installed capacity in 3.7--928.6~s per benchmark, improving on
Manual in four cases and matching it in three. On Sapphire Rapids,
\mbox{NoPG-B} groups the cores and LLC slices on two directly connected
dies, doubling the number of slices local to each core from 15 to 30.
The remaining slices are one hop away, and installed package capacity
falls by 42.2\%. On MI300X and MI300A, \mbox{NoPG-B} redistributes
Infinity Cache traffic across inter-die links, supporting the same
traffic domain \(\Omega\) with less installed capacity.
For the seven routed \mbox{NoPG-B} topologies in
Table~\ref{tab:qor-comparison}, independently minimizing each link's
rounded capacity over its interface options gives a lower bound equal
to \(B_{\mathrm{pkg}}\) in four cases and at most 2.02\% below it in the
other three.
To assess how well the continuous
cost approximation (Section~\ref{sec:package-cost}) guides design selection,
we replayed all recorded \mbox{NoPG-C} link configurations without the cost cutoff.
The original search skipped floorplanning for 1,269 of 1,360 configurations
(93.3\%). On every benchmark, the returned design cost at most 0.61\%
more than the cheapest layout found in replay.

\subsubsection{KaHyPar partitioning}
With the same backend, \mbox{NoPG-C} lowers both package cost and
amortized NRE relative to \(\mathrm{NoPG}_{\mathrm{KaHyPar}}\) on all
seven benchmarks, with a 12.9\% reduction in geometric mean package cost.
On GB200~GPU it uses three
designs for 21 logic dies, while KaHyPar uses ten for 11.
The NRE saving accounts for 59\% of the package cost gap at \(Q=10^6\).
\mbox{NoPG-B} uses less capacity on three benchmarks and ties KaHyPar
at the HBM lower bound on GA100 and GH100.
In capacity mode, the KaHyPar-based flow finds no feasible layout
for either MI300 benchmark.
Of the 18 MI300X partitions that pass the initial checks, eight exceed
the cost ceiling on die cost alone and five yield no routed topology
within the search budget. For the remaining five, the package search
finds no feasible configuration within the cost ceiling.
On Sapphire Rapids, KaHyPar's weighted cut is smaller than
\mbox{NoPG-B}'s (9.16 vs.\ 10.17~TB/s).
Equation~\eqref{eq:partition-pair-count} counts 54 communicating die pairs
in KaHyPar's partition and 13 in \mbox{NoPG-B}'s.
With KaHyPar's partition, 74.2\% of core--LLC
pairs use multihop routes that need capacity on every link along the path.
The package installs 27.328~TB/s versus 5.920~TB/s for \mbox{NoPG-B}.

\subsubsection{ChipletPart}
With both tools restricted to the same 2.5D packaging options,
\mbox{NoPG-C} finds a valid layout on all seven benchmarks and
finishes each search sooner than ChipletPart. \mbox{NoPG-C} also has lower
package cost on all five benchmarks where ChipletPart reports a valid layout.
On GA100 and GB200~GPU, ChipletPart returns no design that passes its
own validity check. For a given partition, its input netlist fixes which
dies must connect directly. In floorplans reconstructed from its selected partitions,
some of those links exceed the 2-mm interface reach in both cases.
\mbox{NoPG-C} can adapt the topology to the package geometry by
routing traffic through intermediate dies.
Its longest routes span two hops on GA100 and three on GB200~GPU.

\subsection{Batch Merge Ablation}
\label{sec:batch-merge-ablation}

\begin{figure}[!htb]
  \centering
  \includegraphics[width=\columnwidth]{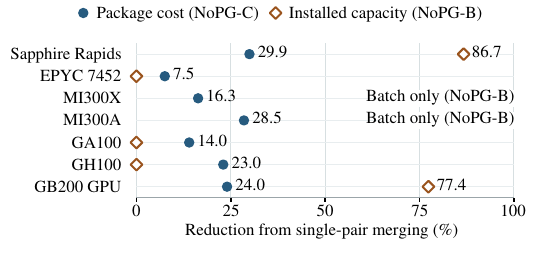}
  \caption{Batch versus single-pair merging at \(Q=10^6\).
  \emph{Batch only} marks capacity-mode cases where only batch merging
  found a feasible layout within the search budget.}
  \label{fig:batch-merge}
\end{figure}

We compare batch merging (Section~\ref{sec:partition-merges}) with
single-pair merging under the same partition-search time budget and
backend settings. Batch merging lowers package cost on all seven
benchmarks, with a 20.8\% reduction in geometric mean cost
(Fig.~\ref{fig:batch-merge}).
In capacity mode, it finds feasible layouts on two additional benchmarks
and lowers geometric mean installed capacity by 50.4\% on the five where
both variants find a feasible layout.

\subsection{Partitioning and Packaging Technology}
\label{sec:packaging-choices}

Joint search finds the lowest package cost on all seven benchmarks
(Table~\ref{tab:packaging-choices}). With the full packaging library
available, repartitioning lowers geometric mean cost by 23.6\% relative
to keeping the reference partition. With repartitioning enabled, expanding
the packaging options beyond the reference technologies lowers it by 12.1\%.
In three of the four restricted MI300 comparisons, our heuristic search
yields designs costing 3.7--7.3\% more than the carefully engineered
Manual designs, but joint search lowers cost on both benchmarks.
For GB200~GPU, \emph{Fixed partition} leaves few packaging choices even with
the full library. The two identical, nearly reticle-sized logic dies must both stay
on the bottom layer for their planar HBM links. Under \emph{Restricted
packaging}, all dies must fit side by side on the Si interposer.
The \mbox{5-nm} blocks (1607~mm\(^2\)) and eight
HBM stacks already occupy 2489 of the available 2905~mm\(^2\), leaving
little room for extra interfaces and layout whitespace. Joint
search keeps the HBM controllers on the interposer and stacks 16
identical \mbox{6-nm} compute dies, reducing cost by 31.9\% relative
to planar repartitioning.

\begin{figure*}[!t]
  \centering
  \includegraphics[width=\textwidth]{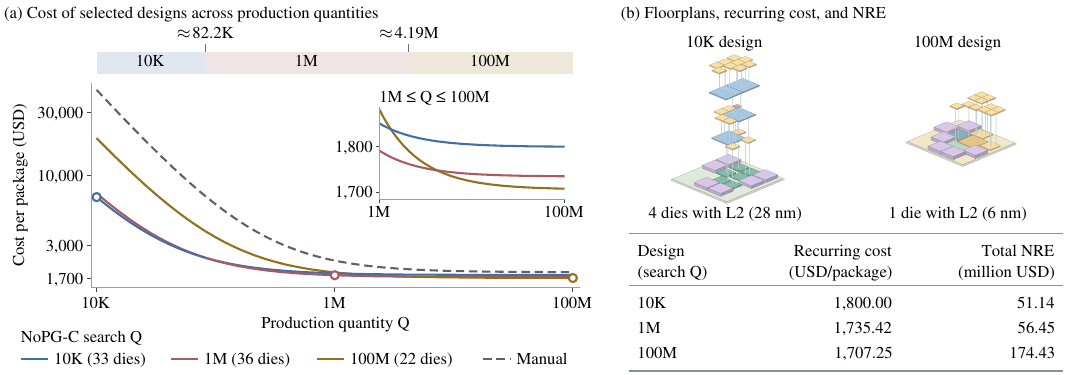}
  \caption{Volume-dependent trade-off between NRE and recurring cost in GA100 package design.}
  \label{fig:cost-results}
\end{figure*}

\subsection{GA100 Case Study}
\label{sec:ga100-case}

Figure~\ref{fig:ga100-objectives} compares the Manual and synthesized
GA100 designs at \(Q=10^6\). \mbox{NoPG-C} needs more link capacity than
\mbox{NoPG-B} for traffic between logic dies, but its smaller, reusable
dies save \$181.12 in fabrication and \$321.43 in amortized NRE per package.
When restricted to planar links on a Si interposer
(\emph{Restricted packaging} in Table~\ref{tab:packaging-choices}),
\mbox{NoPG-C} finds a 17-die design that groups the 12 HBM controllers
in pairs with NVLink interfaces and places L2 slices with SMs. With
stacking available, the 36-die design places each controller pair on a
dedicated 12-nm die and uses 20 vertical
links. Keeping this partition and routing rules out a potentially
lower-cost planar package on an organic substrate. On the 28-nm PCIe
die that forwards traffic, the six links
would need 154.9~mm of die edge with organic interfaces. Even a die as
large as the \(26\times33\)~mm reticle has only 118~mm of perimeter.

Figure~\ref{fig:cost-results} compares Manual with \mbox{NoPG-C} designs
synthesized at 10K, 1M, and 100M units.
We fix each layout and assembly order and amortize NRE over \(Q\)
packages as in~\eqref{eq:stage1-die-ledger}. At each target volume, the
corresponding design has the lowest cost per package.
At 10K units, vertical links
supply the required bandwidth between small \mbox{6-nm} compute dies and
separate \mbox{28-nm} L2 dies. Reusing both designs limits upfront NRE.
At~100M units, recurring cost dominates. The design integrates L2 with
compute on a \mbox{6-nm} die and saves \$15.80 per package in die
fabrication relative to the 10K design. The larger saving is in assembly
(\$79.97 per package), where fewer vertical bonds (12 versus 27)
reduce exposure to yield losses.

\section{Limitations and Future Work}
\label{sec:limitations}

We report the feasible designs returned within the search budget for
the given communication requirements and technology library.
When scaling block area, we assume each functional block can meet the
same performance target on any of its allowed process nodes.
NoPG enforces inter-chiplet bandwidth and hop constraints and estimates
the area of each chiplet's NoC from the traffic it carries, including
the NoP traffic it forwards. We defer end-to-end latency analysis under
contention and verification of deadlock freedom until the chiplet-local
NoCs and inter-chiplet flow control are specified.
Co-locating communicating blocks to reduce
installed capacity \(B_{\mathrm{pkg}}\) avoids the energy cost of
off-die signaling for those transfers. However, explicit power and
thermal modeling falls outside the current framework. For each reusable
die, we check that one set of interface macro sites can serve every
copy's links. Individual bump assignments and detailed PHY layout
remain part of die implementation. We plan to couple process-dependent
models of block performance and power with package thermal analysis,
and to add NoC synthesis and end-to-end latency analysis.

\section{Conclusion}
\label{sec:conclusion}

In this paper, we explore how to choose chiplet boundaries and interconnects
for cost-effective 2.5D and 3D integration. Starting from functional blocks
and their communication requirements, NoPG synthesizes feasible
package designs for a given production volume using the available process
and packaging options.
Experiments on seven benchmarks modeled on commercial chiplet-based designs
show that package-aware co-design can lower package cost
relative to the reference designs or reduce installed link capacity within
their cost ceilings.

\section*{Acknowledgment}
OpenAI GPT-6 was used for language editing, coding assistance, and the development of plotting scripts.
All generated content was reviewed by the authors.

\bibliographystyle{IEEEtran}
\bibliography{references}

\end{document}